\documentclass[]{emulateapj}

\newcommand{\lessim}{\lesssim}
\newcommand{\gas}{\mathrm{gas}}

\newcommand{\SFR}{\mathrm{SFR}}
\newcommand{\LCDM}{\Lambda\mathrm{CDM}}
\newcommand{\OmegaL}{\Omega_{\Lambda}}
\newcommand{\OmegaM}{\Omega_{\mathrm{M}}}
\newcommand{\Omegab}{\Omega_{\mathrm{b}}}

\newcommand{\Mvir}{M_{\mathrm{vir}}}
\newcommand{\Mlow}{M_{\mathrm{low}}}
\newcommand{\zlow}{z_{\mathrm{low}}}
\newcommand{\zhigh}{z_{\mathrm{high}}}
\newcommand\Mvirn[1]{M_{\mathrm{vir,#1}}}

\newcommand{\Mstar}{M_{\star}}
\newcommand{\Mgas}{M_{\gas}}
\newcommand{\Msun}{M_{\sun}}

\newcommand{\sigmae}{\sigma}

\newcommand{\BH}{\mathrm{BH}}

\newcommand{\MBH}{M_{\mathrm{BH}}}

\newcommand{\Mbulge}{M_{\mathrm{bulge}}}

\newcommand{\msigma}{\MBH\mathrm{-}\sigmae}
\newcommand{\mbulge}{\MBH\mathrm{-}\Mbulge}

\newcommand{\fgas}{f_{\gas}}

\newcommand{\AB}{\mathrm{AB}}
\newcommand{\Vega}{\mathrm{Vega}}
\newcommand{\mum}{\mu\mathrm{m}}
\newcommand{\Ang}{\mbox{\AA}}
\newcommand{\fACSB}{\mathrm{F435W}}
\newcommand{\fACSV}{\mathrm{F606W}}
\newcommand{\fACSi}{\mathrm{F775W}}
\newcommand{\fACSI}{\mathrm{F814W}}
\newcommand{\fACSz}{\mathrm{F850LP}}
\newcommand{\fNICMOSJ}{\mathrm{F110W}}
\newcommand{\fNICMOSH}{\mathrm{F160W}}
\newcommand{\fRs}{R_{s}}
\newcommand{\mRs}{R_{s}}
\newcommand{\mG}{G}

\newcommand{\mUn}{U_{n}}

\newcommand{\mKs}{K_{s}}
\newcommand{\mI}{I}
\newcommand{\mY}{Y}
\newcommand{\mJ}{J}
\newcommand{\mH}{H}
\newcommand{\Ks}{K_{s}}
\newcommand{\muprime}{u'}
\newcommand{\mgprime}{g'}
\newcommand{\mrprime}{r'}
\newcommand{\miprime}{i'}
\newcommand{\mzprime}{z'}

\newcommand{\mACSB}{B_{435}}
\newcommand{\mACSV}{V_{606}}
\newcommand{\mACSi}{i_{775}}

\newcommand{\mACSz}{z_{850}}
\newcommand{\mNICMOSJ}{J_{110}}
\newcommand{\mNICMOSH}{H_{160}}
\newcommand{\fIRACa}{3.6\mum}
\newcommand{\fIRACb}{4.5\mum}
\newcommand{\fIRACc}{5.8\mum}
\newcommand{\fIRACd}{8.0\mum}
\newcommand{\mIRACa}{[3.6\mum]}
\newcommand{\mIRACb}{[4.5\mum]}
\newcommand{\mIRACc}{[5.8\mum]}
\newcommand{\mIRACd}{[8.0\mum]}
\newcommand{\AND}{\wedge}
\newcommand{\OR}{\vee}
\newcommand{\dd}{\mathrm{d}}
\newcommand{\SB}{\mathrm{Starburst 99}}

\newcommand{\Gyr}{\mathrm{Gyr}}
\newcommand{\Myr}{\mathrm{Myr}}
\newcommand{\yr}{\mathrm{yr}}
\newcommand{\pc}{\mathrm{pc}}
\newcommand{\kpc}{\mathrm{kpc}}
\newcommand{\Mpc}{\mathrm{Mpc}}
\newcommand{\Gpc}{\mathrm{Gpc}}

\newcommand{\ergs}{\mathrm{ergs}}
\newcommand{\cm}{\mathrm{cm}}
\newcommand{\Jy}{\mathrm{Jy}}

\newcommand{\Lsun}{L_{\sun}}
\newcommand{\Lbol}{L_{\mathrm{bol}}}
\newcommand{\dotMBH}{\dot{M}_{\mathrm{BH}}}

\newcommand{\Lya}{\mathrm{Ly}\alpha}
\newcommand{\alphaOX}{\alpha_\mathrm{OX}}
\newcommand{\NH}{N_{H}}
\newcommand{\AV}{A_{V}}
\newcommand{\RV}{R_{V}}
\newcommand{\Alambda}{A_{\lambda}}

\shorttitle{Massive High-Redshift Galaxies}
\shortauthors{Robertson et al.}

\begin{document}

\title{Photometric Properties of the Most Massive High-Redshift Galaxies}
\author{Brant Robertson\altaffilmark{1,2,4,5},
	Yuexing Li\altaffilmark{3},
	Thomas J. Cox\altaffilmark{3},
	Lars Hernquist\altaffilmark{3} \&
	Philip F. Hopkins\altaffilmark{3}}

\altaffiltext{1}{Kavli Institute for Cosmological Physics, and Department of Astronomy and
Astrophysics, University of Chicago, 933 East 56th Street, Chicago, IL 60637, USA}
\altaffiltext{2}{Enrico Fermi Institute, 5640 South Ellis Avenue, Chicago, IL 60637, USA}
\altaffiltext{3}{Harvard-Smithsonian Center for Astrophysics, 
        60 Garden St., Cambridge, MA 02138, USA}
\altaffiltext{4}{Spitzer Fellow}
\altaffiltext{5}{brant@kicp.uchicago.edu}

\begin{abstract}

We calculate the observable properties of the most massive high-redshift galaxies
in the hierarchical formation scenario where stellar spheroid and supermassive black hole 
growth are fueled by gas-rich mergers.
Combining high-resolution hydrodynamical simulations of the hierarchical formation  
of a $z\sim6$ quasar, stellar population synthesis models, template AGN spectra, 
prescriptions for 
interstellar and intergalactic absorption, and the response of modern telescopes,
the photometric evolution of galaxies destined to host $z\sim6$ quasars are modeled
at redshifts $z\sim4-14$.
These massive galaxies, with enormous stellar masses of $\Mstar \sim 10^{11.5}-10^{12}\Msun$ 
and star formation rates of $\SFR\sim10^{3}-10^{4}\Msun \yr^{-1}$ at $z\gtrsim7$,
satisfy a variety of photometric selection criteria based on Lyman-break techniques
including $V$-band dropouts at $z\gtrsim5$, $i$-band dropouts at $z\gtrsim6$, and
$z$-band dropouts at $z\gtrsim7$.  The observability of the most massive high-redshift
galaxies is assessed and compared with a wide range of existing and proposed
photometric surveys including SDSS, GOODS/HUDF, NOAO WDFS, UKIDSS, the IRAC Shallow
Survey, Ultra-VISTA, DUNE, Pan-STARRS, LSST, and SNAP.  
Massive stellar spheroids descended from $z\sim6$ quasars
will likely be detected at $z\sim4$ by existing surveys, but
owing to their low number densities
the discovery of
quasar progenitor galaxies at $z>7$
will likely require future surveys of large portions of the sky ($\gtrsim0.5\%$)
at wavelengths $\lambda\gtrsim1\mum$.
The detection of
rare, star-bursting, massive galaxies at redshifts $z\gtrsim6$ would provide
support for the hierarchical formation of the earliest
quasars and characterize the primitive star-formation histories of 
the most luminous elliptical galaxies.
\end{abstract}

\keywords{galaxies: formation -- galaxies: evolution}

\section{Introduction}
\label{section:introduction}

Supermassive black holes (SMBHs) serve as the
relativistic engines of quasars
\citep{lynden-bell1969a} and
the existence of quasars at $z\sim6$ 
suggests the rapid formation of SMBHs
with masses $\MBH\sim10^{9}\Msun$
\citep[e.g.,][]{fan2000a,fan2001a,fan2003a,fan2004a,fan2006a}.
Examining the consistency of rapid SMBH growth and the formation of
cold dark matter (CDM) structures at high-redshifts has been 
of long standing theoretical interest \citep[e.g.,][]{efstathiou1988b},
as a confluence of the deep potential well of a rare, massive dark 
matter halo, an abundant supply of baryons, and an efficient SMBH
fueling mechanism is likely needed to grow enormous SMBHs
in the available $\approx 900$ Myrs before redshift $z\sim6$.
The existence of quasars at $z\sim6$ can be considered \it a posteriori \rm
evidence that our currently dark energy-dominated ($\LCDM$) universe 
can muster this confluence, but 
little else is known observationally about the associated formation of quasar host
galaxies at such high redshifts.  The focus of this paper is to 
assist in the detection and characterization of massive
galaxies destined to host $z\sim6$ quasars by 
calculating their photometric evolution in the scenario where
$\LCDM$ structure formation fuels SMBHs and grows stellar
spheroids through gas-rich mergers.

The formation of stellar spheroids through mergers of disk
galaxies has been a long-established vehicle for galaxy evolution
\citep[e.g.,][]{toomre1972a,toomre1977a}.  If the disk galaxies
contain gas before the merger, gravitational torques can enable
gas to loose angular momentum, flow efficiently to the centers
of the merging systems, and result in a starburst 
\citep[][]{larson1978a,noguchi1988a,hernquist1989a,barnes1991a,barnes1996a,mihos1994a,mihos1996a}.
The efficient fueling of starbursts by mergers may
explain ultraluminous infrared galaxies 
\citep[ULIRGS, e.g.][]{soifer1984a,young1986a,soifer1987a,sanders1996a}
as an intermediate phase where the high infrared luminosities are powered
by star formation before gas accretion can grow central  
supermassive black holes to be massive enough to become luminous quasars
\citep[][]{sanders1988a,sanders1988b,hernquist1989a,sanders1996a,genzel1998a}.
This evolutionary picture would paint starbursts, ULIRGS, and quasars as 
stages in a structure formation process that builds spheroid and SMBH
mass contemporaneously through galaxy mergers.

The observed correlations between SMBH mass and stellar spheroid mass 
\citep[the $\mbulge$ relation, e.g.][]{kormendy1995a,magorrian1998a,marconi2003a,haring2004a}, 
SMBH mass and stellar velocity dispersion
\citep[the $\msigma$ relation, ][]{ferrarese2000a,gebhardt2000a,tremaine2002a},
and the black hole fundamental plane \citep{hopkins2007a}
have been widely interpreted as evidence for the coeval or coupled formation of SMBHs
and stellar spheroids 
\citep[e.g.][]{silk1998a,fabian1999a,kauffmann2000a,volonteri2003a,wyithe2003a,king2003a,granato2004a,
hopkins2006a, hopkins2006b}.
Hydrodynamical simulations of the formation of SMBHs and their host spheroids in gas-rich mergers 
\citep{di_matteo2005a,robertson2006a} that include a prescription for SMBH growth and feedback
\citep{springel2005b} generically predict that such mergers undergo starbursts, consistent with 
earlier simulations that did not include SMBH modeling \citep[e.g.,][]{mihos1994a,barnes1996a}, and
that the spheroid forms primarily before the growing SMBHs reach their final mass.
Once the rapidly growing SMBHs become massive enough to drive energetic outflows, the
SMBH feedback truncates star formation \citep[][]{di_matteo2005a,springel2005e} and becomes visible
as a luminous quasar as the outflow removes obscuring material at the galaxy center \citep{hopkins2005a}. 
Ultimately, the end state of this dissipative process is a spheroidal remnant
that closely matches observed elliptical galaxy 
structural \citep[e.g.,][]{robertson2006b} and kinematic \citep[e.g.,][]{cox2006a}
properties and satisfies the observed SMBH-spheroid correlations \citep[][]{di_matteo2005a,robertson2006a,
hopkins2007a}. 
The general consistency of the evolutionary sequence as detailed by these simulations with available 
observational data at redshifts $z\lesssim3$ has been discussed elsewhere 
\citep[e.g.,][]{hopkins2006a, hopkins2006b, hopkins2007b}.

While a developmental connection between stellar spheroids and the SMBHs they
host has observational support in the nearby universe, the properties of
$z\sim6$ quasars may complicate this simple coeval picture.
The Sloan Digital Sky Survey \citep[SDSS;][]{york2000a} was used in concert with
Two Micron All Sky Survey $J$-band imaging 
\citep[2MASS;][]{skrutskie1997a} 
to
photometrically identify the quasar SDSS J1148+5251, spectroscopically confirmed to
reside at redshift $z=6.42$ \citep{fan2003a}.  Estimates of the SMBH mass
from MgII and CIV line widths for this quasar are 
$\MBH \sim 2-6 \times 10^{9}$ \citep{willott2003a,barth2003a}, suggesting that the
SMBH is near its Eddington limit. 

Radio observations of
SDSS J1148+5251 infer the presence of molecular hydrogen from CO(3-2) emission, 
with $M_{\mathrm{H}2} = 2.2 \times 10^{10}\Msun$ \citep{walter2003a}.  The
dynamical mass for the quasar host $M_{\mathrm{dyn}}\sim4.5\times10^{10}\Msun$,
estimated from CO line widths \citep{walter2003a,walter2004a}, suggests that the 
molecular gas and SMBH, rather than some dominant stellar component,
supply the majority of the local potential within $\sim2.5\kpc$.
However, radiative transfer calculations of CO line emission in quasars 
show that owing to projection effects the small observed line widths
could be consistent with presence of a  
massive bulge in the 
quasar host galaxy, but such observed orientations are unlikely \citep{narayanan2006a}.
The large IR luminosity $L_{\mathrm{IR}}\approx10^{13}\Lsun$ \citep{bertoldi2003b}
and radio continuum imaging \citep{carilli2004a} imply
a star formation rate of $\sim3,000\Msun\yr^{-1}$ if the emission is powered
by reprocessed starlight, and the presence
of dust and metals may suggest a prior epoch of star formation in the system
\citep{barth2003a,walter2003a,bertoldi2003b,carilli2004a}.
The general uncertainty about the presence of a massive stellar component in 
$z\sim6$ quasars then poses two questions:
if the progenitors and descendants of $z\sim6$ quasars do contain a substantial
stellar component what
would their properties be and how might one detect them?

The detection of high-redshift galaxy populations at
redshifts $z\gtrsim3$ has greatly developed over the last fifteen years.
The
adoption of photometric techniques to identify the Lyman break in galaxy
spectra \citep[e.g.,][]{steidel1992a,steidel1993a,steidel1996a}, 
owing to intergalactic medium (IGM)
absorption \citep[e.g.,][]{madau1995a}, the Balmer and $4000\Ang$ 
spectral break
\citep[e.g.,][]{franx2003a}, and narrow-band searches for 
$\Lya$ emission \citep[e.g.][]{cowie1998a,hu2002a}
has led to the discovery of
galaxy populations at progressively higher redshifts.
Multicolor \it Hubble Space Telescope \rm(HST) observations of the 
Hubble Deep Field \citep[HDF;][]{williams1996a} allowed for the efficient
identification of large numbers of 
$\mACSB$-band ($\fACSB$ filter) dropout candidates at 
$3.5\lesssim z\lesssim4.5$ \citep{madau1996a}
and redder, more distant dropout candidates \citep[e.g.,][]{lanzetta1996a}
including a $\mACSV$-band ($\fACSV$ filter) dropout spectroscopically
verified to reside at $z=5.34$ \citep{spinrad1998a}.

Subsequently, photometric dropout techniques based on the
Lyman spectral break were used to identify high-redshift
galaxy populations in the Hubble Ultra Deep Field \citep[HUDF;][]{beckwith2006a}
with the Advanced Camera for Surveys \citep[ACS;][]{ford1998a} and
the Near Infrared Camera and Multi-Object Spectrometer (NICMOS)
Ultra Deep Field \citep[][]{thompson2005a}, and in the wider
Great Observatories Origins Deep Survey 
\citep[GOODS;][]{dickinson2003a,giavalisco2004b}.
As with the HDF data,
color selections based
on $\mACSB$- and $\mACSV$-dropouts were used \citep[e.g.,][]{giavalisco2004a}, 
but large samples based on $\mACSi$-band ($\fACSi$ filter) 
and even $\mACSz$-band ($\fACSz$ filter) dropouts have been
identified in GOODS and the HUDF by several groups
\citep[e.g.,][]{stanway2003a,stanway2004a,stanway2004b,stanway2005a,yan2003a,yan2004a,yan2004b,yan2006a,bunker2003a,bunker2004a,bouwens2003a,bouwens2004a,bouwens2006a,bremer2004b,dickinson2004a,stiavelli2004a,eyles2006a,dunlop2006a,overzier2006a,lee2006a,reddy2006a,vanzella2006a,verma2007a}.
The GOODS optical and near-infrared data sets have been combined
with \it Spitzer Space Telescope \rm \citep{werner2004a} observations using 
the Infrared Array Camera \citep[IRAC;][]{fazio2004a} 
to help constrain
longer wavelength portions of the spectral energy distributions 
(SEDs) of high-redshift galaxy candidates \citep[e.g.,][see also \cite{barmby2004a}]{mobasher2005a,egami2005a,yan2005a,eyles2005a,stark2006a,papovich2006a,labbe2005a,labbe2006a}. 

Parallel efforts have been made with ground-based observations,
sometimes in concert with HST or \it Spitzer \rm data sets, including
optical break searches \citep{franx2003a,daddi2003a,van_dokkum2004a},
$z\gtrsim5$ Lyman-break galaxy (LBG) searches using a variety of techniques 
including $RIz$ \citep{lehnert2003a},
$BVRiz$ \citep{ouchi2004a}, $Bz\Ks$ \citep{grazian2006a}, and
$zSZJHK$ \citep{richard2006a} 
color selections, 
and very-high redshift ($z\gtrsim5$) narrow band and spectroscopic $\Lya$-emitter 
searches
\citep[e.g.,][]{ellis2001a,kodaira2003a,kurk2004a,santos2004a,ouchi2005a,chary2005a,taniguchi2005a,stern2005a,iye2006a,kashikawa2006a}.  
Collectively, these observational efforts and others have established the existence of
a substantial galaxy population at $z>5$.  A brief review of 
the known high-redshift galaxy population can be found in \cite{hu2006a}.

Theoretical calculations of the photometric properties of galaxies 
have been explored previously by
hydrodynamical simulations of cosmological structure formation
at
redshifts
$z<3$ \citep{katz1992a,dave1999a,nagamine2005a,nagamine2005b},
$z\sim3$ \citep{weinberg2002a,nagamine2004a,nagamine2006a},
$z\gtrsim4$ \citep{harford2003a,finlator2006a,night2006a},
and $z\gtrsim6$ \citep{barton2004a,dave2006a}, and in
semi-analytic calculations \cite[e.g.,][]{baugh1998a,kauffmann1999a,somerville2001a}.
These calculations show that simulated galaxies, typically with
maximum star formation rates of $\SFR\lessim100\Msun\mathrm{yr}^{-1}$,
can satisfy Lyman-break color selection techniques.  These
simulated galaxy samples broadly agree with the observed
properties of high-redshift galaxies, but given the simulated
comoving volumes ($\lessim0.001h^{-3}\Gpc^3$), the galaxies
examined to date are orders of magnitudes less massive than $z\sim6$
quasar host halos that have comoving number densities of $n\sim1\Gpc^{-3}$ 
\citep{fan2003a}.
Furthermore, these cosmological simulations
have not included prescriptions for energetic feedback from
SMBH growth that have been demonstrated
to influence the photometric properties of 
elliptical remnants formed in
hydrodynamical simulations of isolated disk
galaxy mergers \citep[e.g.,][]{springel2005e}.

To calculate the photometric evolution of $z\sim6$ quasar host galaxies,
we use the simulations of \cite{li2006a} that combine
merger trees measured from
cosmological N-body simulations of a $1h^{-3}\Gpc^3$ comoving
volume with hydrodynamical simulations of galaxy mergers that
include gas cooling, star formation and supernovae, and a prescription
for feedback from growing SMBHs to track the hierarchical formation
of a massive galaxy at redshifts $z\lesssim14$.  These simulations
are described in detail in \cite{li2006a} and summarized in 
\S \ref{section:simulations}.  The galaxy hosts
a quasar at $z\sim6.5$ that is similar in bolometric luminosity to
the $z=6.42$ quasar SDSS J1148+5251 \citep{fan2003a}.  The photometric 
evolution of the host galaxy is modeled before, during, and after
the quasar phase by applying population synthesis models, template AGN
spectra, a
prescription for dust attenuation and reddening, intergalactic medium
absorption, and telescope responses, and is
described in \S \ref{section:photometry}.  In \S \ref{section:evolution},
the photometric evolution
of the
host galaxy is compared with a 
variety of optical and near-infrared
photometric selection criteria from the literature.
The observability of high-redshift quasar host galaxies is assessed in 
\S \ref{section:observability}.
We provide a brief discussion in \S \ref{section:discussion} and summarize 
and conclude in \S \ref{section:conclusions}.
We adopt a flat $\LCDM$ cosmology
with matter density $\OmegaM=0.3$, dark energy density $\OmegaL=0.7$,
baryon density $\Omegab=0.04$, spectral index $n=1$, and root-mean-squared
fluctuations on $8h^{-1}\Mpc$ scales of $\sigma_{8}=0.9$.  Unless
otherwise noted, we report magnitudes on the AB scale \citep{oke1974a}.

\section{Simulations}
\label{section:simulations}
 
We use the \cite{li2006a}
hydrodynamical simulation
of the formation of a $z\sim6$ quasar as input for our
calculation of the photometric evolution of the most massive
quasar host galaxies.
The simulation is described in
detail by \cite{li2006a}, and other specifics of the modeling can be found in
\cite{springel2005b} and \cite{robertson2006a}, but a brief summary of the simulation
follows.

The rarity of $z\sim6$ quasars requires the simulation of
very large volumes to capture the high-sigma
fluctuations that give rise to early-forming massive galaxies.
To accomplish this, a $400^{3}$ particle N-body simulation of
a $1h^{-3}\Gpc^3$ volume of a cosmology consistent 
with the first-year results of the \it Wilkinson Microwave
Anisotropy Probe 
\rm \citep[WMAP1, $\OmegaM=0.3$, $\OmegaL=0.7$, $\Omegab=0.04$, $\sigma_{8}=0.9$,][]{spergel2003a} 
was performed.  
This simulation, with
dark matter particle mass $m_{\mathrm{DM}}=1.3\times10^{12}h^{-1}\Msun$
and force resolution $\epsilon=125h^{-1}\kpc$, was
evolved from $z\sim30$ to $z\sim0$.  Dark matter particles bound to
the most massive
halo in the volume, with $M_{\mathrm{halo}} = 3.6\times10^{15}h^{-1}\Msun$ at $z=0$,
are identified at $z\approx6$.  The $\Gpc^3$ volume is then resimulated \citep{power2003a,gao2005a}
from $z=69$
to $z=0$
with $350^{3}$ dark matter particles, with $340^{3}$ high-resolution
dark matter particles ($M_{dm}=2.8\times10^{8}h^{-1}\Msun$, $\epsilon=5h^{-1}\kpc$) distributed
in the $\sim1.25\times10^{-4}h^{-3}\Gpc^{3}$ Lagrangian volume that 
comprised the
most massive halo at $z\sim6$ and the remaining particles properly placed
throughout the 
rest of the $\Gpc^3$ volume to retain the large scale tidal field.

The $\LCDM$ simulation is used to follow each merger in the 
hierarchical formation history of the halo between $z=14.4$ and $z=6$.
The merger history is comprised of 
7 major mergers (with mass ratios less than $5:1$) between $z=14.4$
and $z=8.5$
contributing to the $5.4\times10^{12}h^{-1}\Msun$ halo mass at $z\sim6$.
To model the formation of the baryonic component of the quasar host halo, each
event in the
merger tree
is resimulated using the smoothed particle hydrodynamics (SPH) / N-body code
GADGET2 \citep{springel2005c}.
The merging systems are represented with hydrodynamical models of disk galaxies 
\citep{springel2005b}
with structural properties initialized according to the \cite{mo1998a}
dissipational galaxy formation formalism and scaled appropriately to
high-redshift following \cite{robertson2006a}.  The scalings simply
fold the redshift- and mass-dependent dark matter concentrations 
\citep[e.g.][]{bullock2001a} and the redshift-dependent Hubble
parameter $H(z)$ through the \cite{mo1998a} formalism, resulting in smaller,
more centrally concentrated disks at higher redshifts.  
The orbital parameters
of each progenitor galaxy (see Table 1 of \cite{li2006a}) are set such that 
the initial separation of the galaxies 
is the current host halo virial radius at the time of the merger and the pericentric
passage distance is half the progenitor disk scale-length $R_{\mathrm{peri}}=0.5R_{\mathrm{d}}$.

The merging disk galaxies range in virial mass from the
initial progenitor's $\Mvir=6.3\times10^{10}h^{-1}\Msun$ at $z=14.4$
to the last merging galaxy's $\Mvir=2.08\times10^{12}h^{-1}\Msun$.
Each progenitor disk has a mass $m_{\mathrm{d}}=0.15\Mvir$, with
gas fractions of $\fgas=1$ at redshifts $z\geq10$ and $\fgas=0.9$
at redshifts $8\leq z<10$.
For the numerical parameters of the progenitor galaxy models, 
a fixed dark matter particle mass
($M_{\mathrm{dm}}=1.1\times10^{7}h^{-1}\Msun$) and baryonic particle mass
($M_{\mathrm{gas}}=M_{\mathrm{star}}=2.2\times10^{6}h^{-1}\Msun$) is used.
The gravitational softening for the dark matter particles 
($\epsilon_{\mathrm{dm}}=60h^{-1}\pc$) is set twice as large as for the 
baryonic particles ($\epsilon_{\mathrm{gas,stars}}=30h^{-1}\pc$).  Star 
formation is treated in a \cite{mckee1977a} multiphase
picture for the interstellar medium (ISM), implemented as a numerical algorithm
by \cite{springel2003a}.
In this multiphase ISM model, dense, star-forming gas particles are treated as a
collection of cold molecular clouds embedded in a tenuous phase heated by supernovae.
The star-formation rate is calibrated to reproduce the Schmidt-Kennicutt Law \citep{schmidt1959a,kennicutt1998a}, with the star forming gas mass calculated from the 
cold phase density.  
Stellar particles are spawned probabilistically from star-
forming gas particles.
Each stellar population is formed with half the mass of
its parent gas particle
and recycles 2\% of its mass 
as metals through supernovae \citep[for details, see][]{springel2003a}.

The \cite{li2006a} simulation includes a prescription for 
SMBH accretion and feedback \citep[for details, see ][]{springel2005b}.  
Each simulated galaxy is initialized with a SMBH ``sink'' particle that is allowed to
spherically accrete through a
\cite{bondi1952a} prescription.  The accretion rate is determined from the
gas temperature and density near the SMBH particle, and is limited to the Eddington
rate.
A fraction $\epsilon_{r}=0.1$
of the accreted rest mass energy released as radiation and a small fraction
of this radiated energy (5\%) is coupled to the surrounding gas as 
thermal feedback.  The strength of the feedback is calibrated to
reproduce the local $\msigma$ relation in isolated mergers of disk
galaxies \citep{di_matteo2005a}.  The mass of the SMBH seed in each
galaxy is determined by assuming a single $200\Msun$ black hole formed
in each halo at $z=30$, in accord with expectations from Population
III star formation \citep[e.g.][]{abel2002a,bromm2004a,yoshida2003a,yoshida2006a}, and then
experienced Eddington-limited growth until its host galaxy
merged into the
quasar host halo.  The SMBH seeds initialized in this manner
range in mass from 
$\MBH=1.5\times10^{4}h^{-1}\Msun$ at $z=14.4$ to 
$\MBH=8.92\times10^{6}h^{-1}\Msun$ at $z=8.5$.  As dynamical 
friction operates on the merging galaxies during the simulation,
the central regions of each system eventually coalesce.  
Once the 
SMBH particles approach to within the gravitational force resolution they
are merged.

\subsection{Mergers, Star Formation Rate, and Active Galactic Nuclei Luminosity}
\label{section:sfr_and_lbol}

\begin{figure}
\figurenum{1}
\epsscale{1}
\plotone{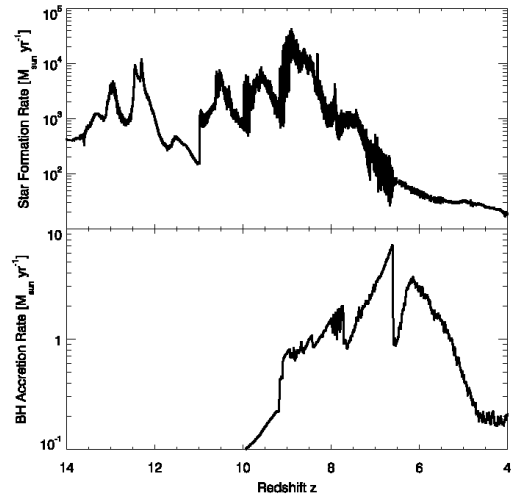}
\caption{\label{fig:bhar_and_sfr}
\small  Star formation rate (SFR) and supermassive black hole (SMBH) accretion rate as a function of redshift in the \cite{li2006a} simulation of the hierarchical formation of a $z\sim6$ quasar.  The upper panel shows the SFR determined from the density of cold
gas over the course of eight galaxy mergers in the build-up to quasar activity at $z\sim6.5$.  The lower panel shows the accretion rate $\dotMBH$ onto SMBHs in the system, determined through a \cite{bondi1952a} spherical accretion model that depends on the properties of gas near the SMBH particles \citep[for details, see][]{springel2005b}.  The unattenuated bolometric luminosity of the active galactic nuclei (AGN) fueled by SMBH accretion can be calculated from the
accretion rate $\dotMBH$.
Adapted from \cite{li2006a}.
}
\end{figure}

The seven major mergers between eight galaxies occur during 
five redshift eras of the evolution:  the first merger
between two galaxies ($\Mvirn{1}=5.3\times10^{10}h^{-1}\Msun$, $\Mvirn{2}=6.3\times10^{10}h^{-1}\Msun$) begins at $z=14.4$, followed by two
additional mergers ($\Mvirn{3}=1.5\times10^{11}h^{-1}\Msun$, $\Mvirn{4}=1.77\times10^{11}h^{-1}\Msun$) at 
$13.0>z>10.5$, single mergers at $10.5>z>9.4$ ($\Mvirn{5}=4.91\times10^{11}h^{-1}\Msun$) and at $9.4>z>8.5$ ($\Mvirn{6}=7.96\times10^{11}h^{-1}\Msun$), and two final 
mergers ($\Mvirn{7}=1.6\times10^{12}h^{-1}\Msun$, $\Mvirn{8}=2.08\times10^{12}h^{-1}\Msun$) at $z\lessim8.5$.  
The fresh gas supply introduced by each merger, combined with
the high density environment and violent tidal interactions, leads to peaks in the
star formation rate (SFR, Fig. \ref{fig:bhar_and_sfr}, upper panel) that temporally 
correlate with the merger events.  
The SFR peaks near the $z\sim8.5$
mergers with $\SFR>10^{4}\Msun \yr^{-1}$, but frequently exceeds $10^{3}\Msun \yr^{-1}$ 
over the redshift range $8<z<13$.  

\normalsize

The total SMBH accretion rate $\dotMBH$ (Fig. \ref{fig:bhar_and_sfr}, lower panel) 
also influences the observable properties of the system.  The unattenuated bolometric 
luminosity $\Lbol$ of the active galactic nuclei (AGN) powered by the growing SMBHs is 
determined by the accretion rate onto the SMBHs through the equation
\begin{equation}
\label{equation:lbol}
\Lbol = \epsilon_{r}\dotMBH c^{2} = 1.45 \times 10^{12} \left(\frac{\epsilon_{r}}{0.1}\right)\left(\frac{\dotMBH}{\Msun \yr^{-1}}\right) \Lsun
\end{equation}
\noindent
where $\epsilon_{r}=0.1$ is the radiative efficiency of the accretion model we adopt and $c$ is the speed of light.

\section{Photometric Modeling}
\label{section:photometry}

Modeling the observed spectral energy distribution (SED) of the quasar host galaxy
involves calculating the SEDs of each individual stellar population and AGN
present in the system 
and applying a wavelength-dependent
dust correction to the calculated SEDs that accounts for internal interstellar
absorption.  
The emission from the galaxy, attenuated by ISM reddening, must then be
corrected for both the IGM opacity owing to Lyman series absorption and redshifting of
the SED owing to cosmological expansion.  The measured photometry of the
object is then modified
by the combined wavelength-dependent transmission of the filter and telescope
assembly as well as the wavelength-dependent quantum efficiency of the detector.
Below we describe how we model each contribution to the observable photometric
properties of the massive quasar host galaxy. 

\subsection{Stellar Population Synthesis}
\label{section:stellar_population_synthesis}

The hydrodynamical simulation follows the formation of individual
stellar populations formed from dense gas, as described in \S \ref{section:simulations}.
Each stellar particle represents several million solar masses of stars, so an
initial mass function (IMF) must be selected to represent distribution of stellar 
masses formed when each stellar particle is spawned.  The IMF also influences the
spectral evolution of each stellar population through its time-dependent
mass-to-light ratio.  We adopt the \cite{kroupa2001a,kroupa2002a} IMF, which
describes the number $N$ of stars of a given mass $m$ as a power-law $\dd N/\dd m \propto m^{-s}$.
The Kroupa IMF shares the same power-law index $s\approx2.3$ as the \cite{salpeter1955a} IMF for 
$m\geq0.5\Msun$ but flattens to $s=1.3$ at lower masses.  The mass-to-light ratio of the Kroupa IMF is
correspondingly lower than for the \cite{salpeter1955a} IMF, by a factor of $\sim2$, when
normalized over the mass range $0.1\Msun\leq m\leq100\Msun$.

The metallicity and age of each stellar particle is tracked in the simulation, allowing
for the use of stellar population synthesis models to calculate an evolving SED for each particle.  
We use $\SB$ \citep{leitherer1999a,vazquez2005a} to model 
the spectra from each population, given the
mass, age, metallicity, and assumed Kroupa IMF.  
The version of $\SB$ we use includes the Padova stellar evolution tracks 
\citep[e.g.,][]{fagotto1994a} to improve the
performance of $\SB$ for older ($t>0.1\Gyr$) stellar populations \citep[for details, see][]{vazquez2005a}.
Illustrations of the time-dependent SEDs produced by $\SB$ can be found in Figures 4 and 5 of 
\cite{vazquez2005a}.
In practice, we produce a grid of $\SB$
spectral models at wavelengths $0.0091\micron\leq \lambda\leq 160\micron$
for stellar populations over the range of ages $t=10^{4}-10^{10}\yr$, spaced 
logarithmically, and for metallicities of $Z=0.02-2.5 Z_{\sun}$, and then interpolate
the $\SB$ models to produce an SED appropriate for each stellar particle.

\subsection{Active Galactic Nuclei SEDs}
\label{section:agn_seds}

As discussed in \S \ref{section:simulations}, the simulations follow accretion
onto SMBH particles.  Given the radiative efficiency we adopt ($\epsilon_{r}=0.1$),
we can calculate the bolometric luminosity of the AGN associated with each 
SMBH (Eq.~\ref{equation:lbol}).  According to the feedback prescription
by \cite{springel2005b}, a fraction $\eta=0.05$ of this bolometric
luminosity is injected as thermal energy in the surrounding gas.  The remaining
$\epsilon_{r}(1-\eta)$ fraction of the radiated
energy would be visible as nuclear emission
in the absence of intervening material.  To model this emission, we combine the
power-law + x-ray reflection component SED template for AGN from \cite{marconi2004a}
with the average quasar emission template from \cite{vanden_berk2001a}.

The \cite{marconi2004a} SED template consists of a power-law $L_{\nu}\propto\nu^{\alpha}$,
with $\alpha=-0.44$ in the range $0.13\micron < \lambda < 1\micron$ set to match the 
slope inferred from stacked SDSS quasar spectra \citep{vanden_berk2001a}, and a 
Rayleigh-Jeans fall-off ($\alpha=2$) at longer wavelengths.  The ultraviolet 
($0.05\micron<\lambda<0.12\micron$) emission has a power-law slope ($s=-1.76$) determined by \cite{telfer2002a}
based on the
HST spectra of $\sim200$ quasars, while the x-ray emission at energies above 1 keV 
is modeled with a power-law plus PEXRAV model \citep{magdziarz1995a} reflection 
component.  The normalization of the x-ray component relative to the optical power-law
component is fixed to reproduce the quantity $\alphaOX$ \citep{zamorani1981a}
that relates the ultraviolet ($\lambda\approx0.25\micron$) to x-ray ($E\approx2$ 
keV) flux ratio.  The \cite{marconi2004a} SED template uses a luminosity-dependent 
$\alphaOX$ following \cite{vignali2003a} that leads to a decreasing fraction of the bolometric
luminosity emitted in x-rays as the AGN luminosity increases.  The ratio of bolometric
luminosity to $B$-band luminosity $\Lbol/\nu_{B}L_{\nu,B}$ has a corresponding 
decrease with bolometric luminosity.  Figure 3 of \cite{marconi2004a} presents
illustrations of their AGN 
SED template and luminosity-dependent $B$-band and x-ray bolometric corrections.

\cite{vanden_berk2001a} produced a composite quasar spectrum by stacking the
spectra of more than 2000 SDSS quasars at wavelengths $0.08\micron<\lambda<8555\micron$, 
including systems with $\mrprime$ absolute magnitudes 
of $-18.0>\mrprime>-26.5$ over the redshift span $0.044\leq z\leq 4.789$.
The continuum of the geometric mean
spectrum of these quasars can be represented by a broken power-law $L_{\nu}\propto \nu^{s}$ with 
$s=-0.44$ over the wavelength range $0.13\micron<\lambda<0.5\micron$ and $s=-2.45$
at longer wavelengths, and is 
representative of the composite spectrum of $z\sim6$ quasars \citep{fan2004a}.
While the short wavelength slope is flatter than the
slope $s\approx-0.9$ found for high-redshift quasars \citep{schneider1991a,fan2001b,schneider2001a}, 
the difference can be accounted for by difficulties in determining the
continuum. 
Prominent emission lines in the \cite{vanden_berk2001a} template spectrum include
$\Lya$, CIV, MgII, and H$\alpha$.

To combine the \cite{marconi2004a} SED template with the \cite{vanden_berk2001a} 
quasar spectrum, we replace the optical power-law component of the \cite{marconi2004a}
with the \cite{vanden_berk2001a} spectrum over 
wavelengths $0.11\micron<\lambda<0.86\micron$ and match the Rayleigh-Jeans and
UV power-law continua to the emission line spectrum.  The spectrum is then integrated
and renormalized to maintain the desired bolometric luminosity, 
and approximately maintains the emission line equivalent
widths and \cite{marconi2004a} luminosity-dependent bolometric corrections while 
scaling the emission line flux with the AGN luminosity.  As discussed below, the 
choice to combine these spectral templates only influences our results during the
evolutionary phases when the AGN dominates the emission of the system, over small
redshift ranges near $z\approx6.5$ and $z\approx6$ \citep[see Fig. \ref{fig:bhar_and_sfr} 
above and Fig. 11 of ][]{li2006a}, but allows us to incorporate the effect of 
$\Lya$ emission on quasar color selection \citep[e.g.,][]{fan2001a}.

\subsection{Interstellar Reddening and Absorption}
\label{section:ism_absorption}

The methods for modeling the emission from stellar populations 
(\S \ref{section:stellar_population_synthesis}) and AGN (\S \ref{section:agn_seds})
allow for an estimation of the unattenuated SEDs of simulated galaxies.  The
gas-rich galaxies involved in the mergers that lead to quasar activity at $z\sim6$
contain substantial columns of hydrogen and, owing to the efficient star 
formation in dense gas, large amounts of obscuring dust.  The wavelength-dependent
attenuation from dust reddening should therefore be accounted for when modeling the
observed SED of the quasar host galaxy.

To incorporate dust attenuation in the photometric modeling, the hydrogen column 
density $\NH$ and an estimate of the gas metallicity is calculated along the 
line-of-sight to each stellar population.  We use the method of \cite{hopkins2005a}
to calculate the gas density on a fine grid from the SPH formalism and the particle
distribution.  For dense gas above the threshold for star formation a 
two-phase decomposition of the \cite{springel2003a} multiphase ISM model is performed,
assuming the hot and cold phases are in pressure equilibrium, in order to calculate the
hot phase column density.  For each AGN in the system the column density and 
metallicity 
is measured along 100 lines-of-sight to calculate a distribution of 
optical depths,  allowing for an estimate of the scatter in the internal galactic absorption.
For discussions on general agreement between the
bound-free and metal-line absorption calculated from this
method for estimating $\NH$ columns to AGN and models of the cosmic x-ray background,
see, e.g., \cite{hopkins2006a,hopkins2007c}.

Once the $\NH$ columns and gas metallicities are calculated, the visual attenuation
 $\AV$ owing to dust is calculated by scaling the Milky Way (MW) dust-to-gas ratio
$\AV/\NH = 5.35\times10^{-22}$ \citep{bohlin1978a,cox2000a} by the gas 
metallicity.  A linear scaling with metallicity, normalized to the MW
ratio for solar metallicity gas, roughly accounts for the observed dust-to-gas
ratios in the less metal-rich Large Magellanic Cloud (LMC) and 
Small Magellanic Cloud (SMC) \citep[e.g.,][]{fitzpatrick1986a,rodrigues1997a}.  
The wavelength-dependent dust attenuation $\Alambda/\AV$ must also be
modeled, with possible choices including the MW dust curve 
\citep[e.g.,][]{cardelli1989a} that includes the strong $0.22\micron$ 
absorption feature, LMC and SMC dust \citep[e.g.,][]{pei1992a,gordon2003a}, and the 
possibly supernovae-related dust in high-redshift quasars \citep{maiolino2004a}.
We adopt the dust model advocated by \cite{calzetti1994a} for modeling the 
UV emission from starburst galaxies, with the updated $\RV=4.05$ from \cite{calzetti2000a} \citep[see also][]{calzetti1997a}.  
The \cite{calzetti2000a} dust law is frequently used to model LBGs and infer their
stellar masses \citep[e.g.,][]{papovich2001a,shapley2005a}, and estimates of 
unattenuated star formation rates using \cite{calzetti2000a} dust corrections
have been used to successfully predict the x-ray and radio properties of galaxies 
at $z\sim2$ \citep{reddy2004a}.  Using SMC dust would increase the UV absorption of
the model spectra and increase the transition redshift where model spectra
cross boundaries in color-color
selections designed to be sensitive to the Lyman break, but would not change the
photometric selection of the model spectra with a given color criterion 
over most of the system's redshift evolution.

\subsection{Intergalactic Medium Absorption}
\label{section:igm_absorption}

Intervening hydrogen clouds in the intergalactic medium (IGM)
can greatly attenuate flux via Lyman-series absorption.  To account for
this IGM opacity, we adopt the \cite{madau1995a} model of
Poisson-distributed absorbers in which
Lyman series blanketing
causes a ``staircase'' absorption profile blue-ward of $\Lya$.
The spectral break introduced by this absorption has
been frequently used to engineer the photometric selection
of high-redshift
galaxies \citep[e.g.,][]{madau1996a,steidel1996a,giavalisco2004a}.
By $z\approx4.5$,
the mean transmission is only 30\% near $\Lya$, but the Poissonian
nature of the model can lead to substantial variation in the
optical depth $\tau$ at intermediate redshifts.
At $z\sim3.5$ the $\pm1\sigma$ fluctuation in optical depth 
corresponds to 30\% transmission variation relative 
to the unattenuated spectra \citep[see, e.g., Fig. 3 of ][]{madau1995a}.
However, Monte Carlo simulations suggest that the bandpass-averaged 
fluctuations in transmission are typically smaller 
\citep[$\sim0.1$ magnitudes, see, e.g., ][]{bershady1999a}.
At high redshifts ($z\gtrsim6$) the absorption is very 
strong near $\Lya$ \citep[e.g., $\tau>3$, see Fig. 2 of ][]{night2006a}
and typical line-of-sight fluctuations will lead to strong spectral breaks. 
We therefore assume the mean IGM opacity calculated from the model since
we are primarily concerned with high-redshifts ($z\gtrsim5$), but we note
that the calculated photometric trajectories could experience either small shifts
in color-color space position or rate of change with redshift owing to
line-of-sight opacity fluctuations.

\subsection{Telescope, Filter, and Detector Response}
\label{section:telescope_response}

To calculate the observed photometric properties of simulated galaxies, the
wavelength-dependent transmission of telescope assemblies, filters,
and detector quantum efficiency (QE) must be modeled together to determine
the net transmissivity of a given observational facility.  

For the SDSS $\muprime$-, $\mgprime$-, $\mrprime$-, $\miprime$-, and $\mzprime$-band 
\citep{fukugita1996a} transmissivities, we use the USNO40 response functions measured by Jim Gunn~\footnote{http://www-star.fnal.gov/ugriz/Filters/response.html}.
The calculated $U_{n}$, $G$, and $\fRs$ \citep{steidel1993a} magnitudes
use the transmissivity measured with the KPNO Mosaic 
CCDs~\footnote{http://www.noao.edu/kpno/mosaic/filters/filters.html}.
For comparison with HST observations, we calculated the wavelength response 
curves from the combined transmissivity of the mirror and filters with the
detector QE for $\fACSB$, $\fACSV$, $\fACSi$, $\fACSI$, and $\fACSz$ observations
with ACS~\footnote{ftp://ftp.stsci.edu/cdbs/cdbs2/comp/acs/} and $\fNICMOSJ$ and $\fNICMOSH$ observations with NICMOS~\footnote{ftp://ftp.stsci.edu/cdbs/cdbs2/comp/nicmos/}.
For the ground-based $\mJ$-, $\mH$-, and $\mKs$-band transmissivities, we use the
calculated relative response curves for 2MASS from \cite{cohen2003a}.
The IRAC throughput response was adopted from data provided by the Spitzer Science 
Center~\footnote{http://ssc.spitzer.caltech.edu/irac/spectral\_response.html}.  
In practice, over the redshift range ($z\gtrsim4$) considered at flux in 
all filters bluer than $\fACSB$ were always too heavily absorbed to be 
easily measured ($>30$ AB magnitudes), so in what follows we focus on
the observable properties of galaxies in redder filters. 

\begin{figure}
\figurenum{2}
\epsscale{1}
\plotone{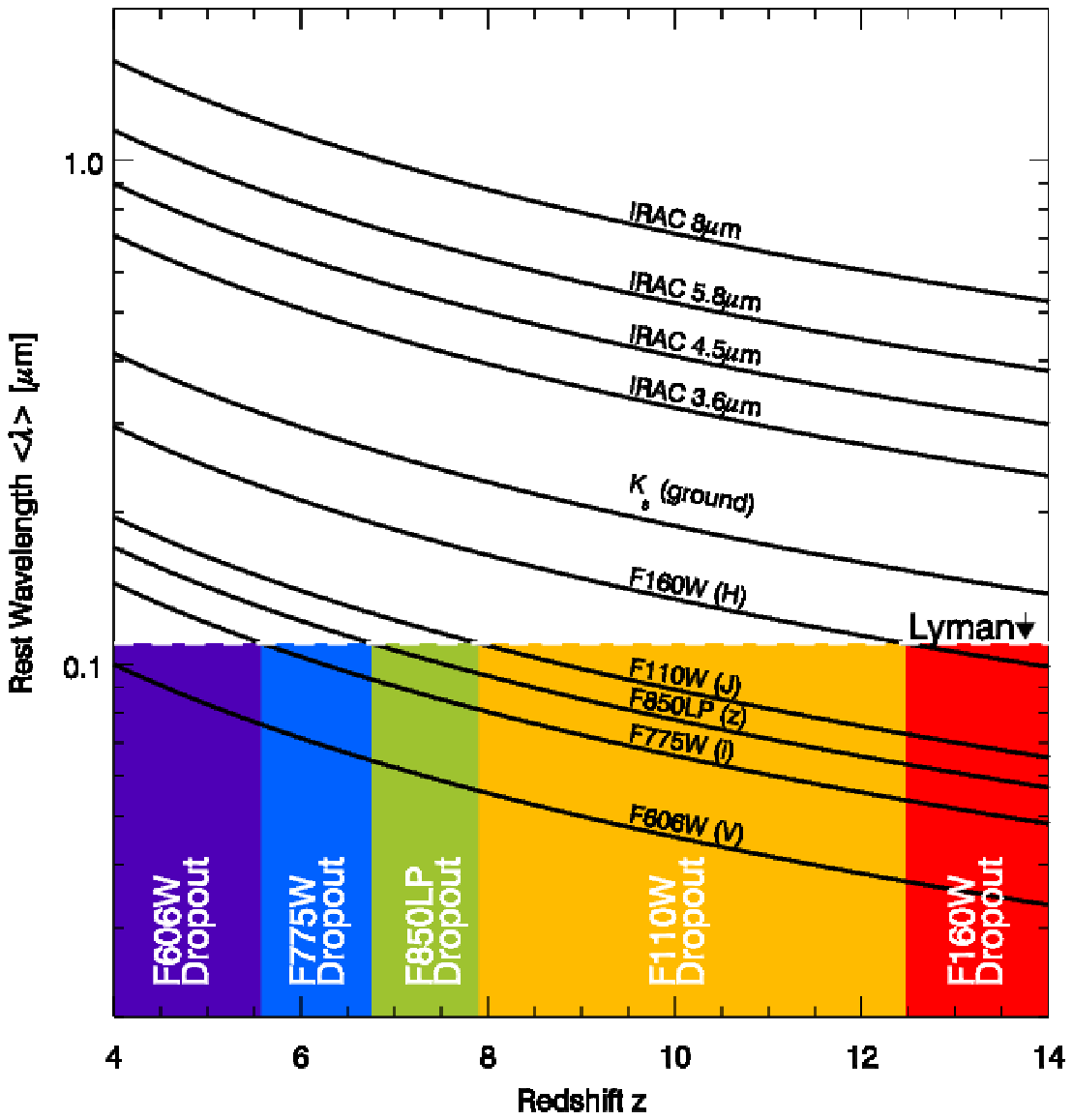}
\caption{\label{fig:dropouts}
\small Schematic diagram of $\Lya$-break dropouts in optical and near-infrared
filters ($\fACSV$, $\fACSi$, $\fACSz$, $\fNICMOSJ$, $\fNICMOSH$, $\Ks$, $\fIRACa$, $\fIRACb$, $\fIRACc$, and $\fIRACd$) at redshifts $z\sim4-14$.  Cosmological expansion 
enables
intergalactic medium (IGM) absorption owing to hydrogen
Lyman-series opacity \citep[e.g.,][]{madau1995a} to absorb light at progressively 
longer observed wavelengths.
As the Lyman-series absorption ($\sim1100\Ang$, white dashed line) shifts into redder 
filters, galaxies begin to drop out of the corresponding photometric samples 
(shaded regions 
represent a $>20\%$ flux decrement in a given band for a flat spectrum).
The solid lines show the effect of cosmological redshifting on the rest-frame
wavelength coverage of commonly used filters.
}
\end{figure}

\subsection{Combined SED Modeling}
\label{section:sed_calculation}

The final model SEDs calculated for the simulated quasar host galaxy
combines the stellar population synthesis models described in 
\S \ref{section:stellar_population_synthesis}, the AGN spectral
template described in \S \ref{section:agn_seds}, ISM attenuation
described in \S \ref{section:ism_absorption}, and IGM attenuation
described in \S \ref{section:igm_absorption}, to arrive at the
observed flux density \citep[adapting the notation of ][]{papovich2001a}
\begin{eqnarray}
\label{equation:flux_density}
F_{\nu}(z, \lambda) &=& (1+z)\frac{\sum_{i}^{N} L_{\nu_{0},i}(\lambda_{0}, t_{i}, A_{\lambda_{0},i}, M_{\star,i}, \dot{M}_{\BH,i})}{4\pi D_{L}^{2}(z)}\nonumber\\
&&\times e^{-\tau_{\mathrm{IGM}}(z,\lambda)},
\end{eqnarray}
\noindent
where the rest-frame emission $L_{\nu_{0},i}$ as a function of rest-wavelength 
$\lambda_{0}$ from each of $N$ sources may depend
on the its stellar age $t_{i}$, stellar mass $M_{\star,i}$, or SMBH accretion rate 
$\dot{M}_{\BH,i}$, and the dust attenuation $A_{\lambda_{0},i}$ along the line-of-sight
to each source.  The typical number of individual stellar population 
sources is $N\gtrsim750,000$ by $z\sim6$.
The total emitted
spectrum from the galaxy is attenuated by the IGM optical depth $\tau_{IGM}$, 
and the factors of $(1+z)$ and the luminosity distance
$D_{L}(z)$ account for the cosmology-dependent redshift and distance.
Again following \cite{papovich2001a}, the flux density in each 
filter is calculated as
\begin{equation}
\label{equation:filter_flux_density}
\left<F_{\nu}(z)\right> = \frac{\int T_{\nu} F_{\nu}(z,\lambda)\dd \nu/\nu}{\int T_{\nu} \dd \nu/\nu}
\end{equation}
\noindent
where unit-free $T_{\nu}$ represents the combined transmissivity of the telescope 
and filter and 
the detector QE, as described in \ref{section:telescope_response}.  The flux
density is converted into a magnitude on the AB scale \citep{oke1974a}
through the relation
\begin{equation}
m_{\AB} = -2.5\log\left(\frac{\left<F_{\nu}\right>}{1\mu \Jy}\right) + 23.9,
\end{equation}
\noindent
where $\left<F_{\nu}(z)\right>$ is the bandpass-weighted flux density (Eq. \ref{equation:filter_flux_density})
and $1\mu \Jy = 10^{-29} \ergs~\cm^{-2} \mathrm{s}^{-1} \mathrm{Hz}^{-1}$.

The machinery described above allows for a thorough determination of the
photometric properties of a simulated galaxy and an application of 
common color ``dropout'' selection techniques based on spectral breaks to track
the photometric evolution of the quasar host galaxy with redshift.
For reference, a schematic diagram of the relevant redshifts for
dropout samples in the ACS and NICMOS filters and a characteristic
wavelength coverage for optical and near-IR filters with redshift is provided in 
Figure \ref{fig:dropouts}.

\begin{figure*}
\figurenum{3}
\epsscale{1}
\plotone{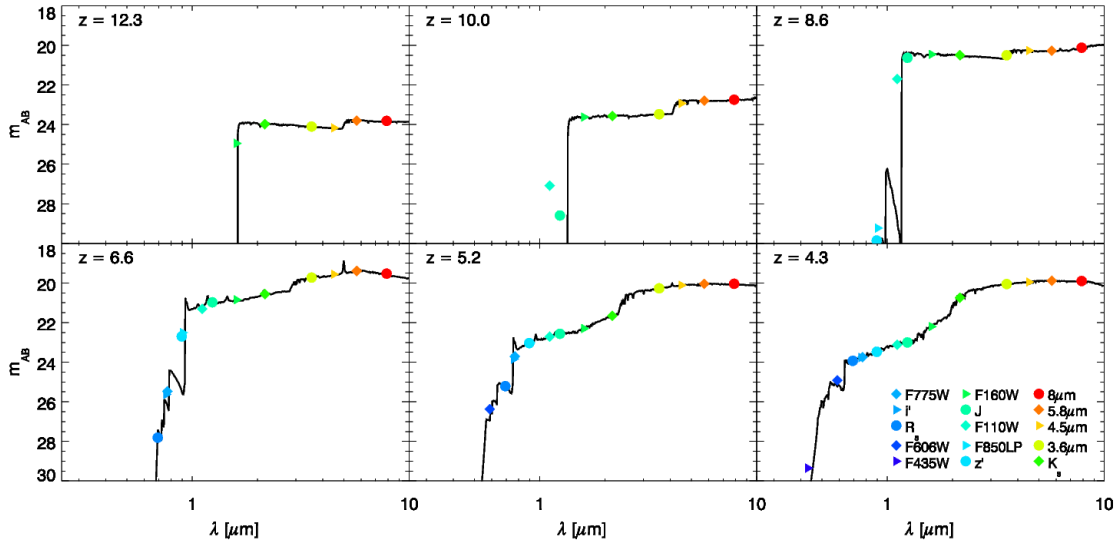}
\caption{\label{fig:seds}
\small Photometric evolution of the quasar host galaxy with redshift.  The spectral 
energy distributions (SEDs), plotted in the observed frame, are 
calculated from stellar and active galactic nuclei
template spectra, attenuated for interstellar absorption, Lyman-series 
absorption
in the intergalactic medium, and cosmological redshifting and distance (see \S \ref{section:photometry}).
At early times ($z\sim8.5-12$, upper panels), the galaxy is rapidly star-forming and 
has ultraviolet-bright
rest-frame spectra typical of starburst galaxies.  By $z\sim6.5$ (bottom left panel),
the SMBH accretion
has peaked, leading to a luminous quasar phase that satisfies the
Sloan Digital Sky Survey
quasar color selection.  As the SMBH accretion drops and the massive bulge ages
($z\lessim6$, bottom middle and right panels), the quasar host galaxy becomes a
progressively redder luminous galaxy with a large Balmer/$4000\Ang$ break 
indicative of an aging stellar
population.  The large bandwidth of some filters can cause an overestimate of the
true flux density when they straddle the $\Lya$ absorption line (e.g., $\fNICMOSJ$ at $z\sim10$).
}
\end{figure*}       

\section{Photometric Evolution}
\label{section:evolution}

The photometric evolution of the quasar host galaxy is calculated
using the methods described in \S \ref{section:photometry}.
In our modeling, we track the luminosity of only the main host halo of the quasar 
during its evolution; the luminosities of the merging systems are not followed 
before they enter the host halo virial radius.  The merging systems are modeled as
almost completely gaseous, resulting in a smooth photometric evolution as each
new system is added to the simulation.  The variations in the system brightness
owe to increases in the star formation rate during the mergers and the 
evolution of the combined stellar populations of the quasar host galaxy.  Below,
we describe the photometric evolution of the simulated system over the
redshifts $z=4-14$.

Figure \ref{fig:seds} shows the SED and 15 magnitudes calculated for the 
galaxy at six characteristic times 
($z\sim12$, $z\sim10$, $z\sim8.5$, $z\sim6.5$, $z\sim5$, and $z\sim4$) 
during three distinct phases in its evolution.
At redshifts $z\gtrsim7$, the system rapidly forms stars as the progenitors of the quasar
host galaxy merge into the system.  
Each of these gas-rich mergers adds enough
fuel to induce large star formation rates of $\SFR\gtrsim10^{3}\Msun\yr^{-1}$.  
This starburst phase coincides with the formation epoch of the stars
that later form the quasar host galaxy.  
Over
this timespan the stellar mass in the system has grown to 
$\Mstar=1.595\times 10^{12}\Msun$, or $96\%$ of the final stellar mass 
at $z=4$, with a mass-weighted mean stellar
age that increases from $\left< t\right>=124\Myr$ at $z=9$ to
$\left< t\right>=185\Myr$
at $z=7.5$.
The upper panels of Figure \ref{fig:seds}
show the characteristic starburst spectrum of the galaxy during this phase, with 
strong rest-frame UV and blue emission.  The SED evolves most strongly at long 
rest-frame wavelengths as the mean stellar age of the system increases, but the
short wavelength rest-frame emission is remarkably similar during this phase.
The luminosity of the system increases as the star formation rate and stellar
mass both increase to $z\sim8$.
The large optical depth owing to 
Lyman-series absorption causes a near step-function break blue-ward of the $\Lya$ line.

By $z\sim7.5$, the quasar host galaxy begins to change rapidly.  At this time a total
gas mass of $\Mgas = 1.21\times10^{11}\Msun$ remains, which amounts to roughly $7\%$
of the total baryonic mass of the system.  Of this gas, $30\%$ has already been 
shocked, heated by feedback, or is otherwise too diffuse to form stars \citep[e.g.,][]{cox2006b}.  
The remaining
$5\%$ of the star-forming baryons is dense enough to support a large star formation rate
of $\SFR\sim1000\Msun\yr^{-1}$.  However, the starburst epoch is clearly over, 
extinguished primarily by the consumption of the available fuel rather than by
some regulatory feedback mechanism.  As the star formation rate rapidly declines,
the stellar population quickly ages and increases
to a mean age of $\left<t\right>=250\Myr$ by $z\sim7$ and $\left<t\right>=407\Myr$ by 
$z\sim6$.  While the gas supply cannot sustain the formation of a large new population
of stars, it can easily support the exponential growth of SMBHs to 
$\MBH\sim10^{9}\Msun$.  As the violent merging of the system continues to drive the 
remaining dense gas toward the galactic centers a rapid phase of 
Eddington-limited SMBH growth begins and eventually leads to a luminous quasar phase
at $z\sim6.5$ after the SMBHs particles coalesce.  The SED of the galaxy reflects the 
quasar activity, with the hybrid
\cite{vanden_berk2001a}-\cite{marconi2004a} model AGN spectrum dominating the blue
emission of the galaxy at $z\approx6.6$ (Fig. \ref{fig:seds}, lower left panel).
The massive stellar component of the system, with a mean stellar age of 
$\sim100\mathrm{s}\,\,\Myr$, contributes substantially to the emission at rest-frame 
wavelengths $\lambda_{0}>0.3\micron$ and instills the SED with a Balmer/$4000\Ang$
break between the $\Ks$- and $\fIRACa$-bands.  As feedback from the SMBH heats the
gas in the central-most regions of the quasar host galaxy, the fuel source for 
SMBH accretion is depleted and the prominence of the AGN contribution to the SED
declines.  By $z\approx5.2$ (Fig. \ref{fig:seds}, lower middle panel) the massive
old stellar population with mean stellar age $\left<t\right>=588\Myr$ dominates the
spectrum, with residual SMBH accretion providing less-dominant AGN activity.  The 
rest-frame UV
and blue optical portion of the SED becomes increasingly feeble compared with the
observed near-IR spectral region powered by mature stars.

Below $z\sim5$, the host galaxy continues to age and redden during its final, passive 
evolutionary phase.  The star formation rate has declined by approximately three 
orders of magnitude from its peak near redshift $z\sim9$.
By $z\approx4.3$ (Fig. \ref{fig:seds}, lower right panel) the
massive stellar component has reached a mean age of $\left<t\right>=873\Myr$, and
by $z\approx4$ the mean stellar age will increase to $\left<t\right>=1.01\Gyr$.  The
UV and blue optical emission from the galaxy has correspondingly decreased, leaving
a SED with a dominant Balmer/$4000\Ang$ break.  The SMBH accretion rate has 
dropped to $\dot{M}_{\BH}\sim0.2\Msun\yr^{-1}$, bringing the bolometric luminosity of
the AGN activity to $\sim10\%$ of the bolometric luminosity of the stellar spheroid.
If the system were to continue to evolve passively to low redshift, 
the galaxy would continue to redden. 
The giant spheroid would eventually become too red to appear 
in LBG samples, such as those defined by $\mUn-\mG$ \citep{steidel1993a,steidel1995a}
and $U_{\mathrm{300}}-B_{\mathrm{450}}$ \citep{madau1996a,steidel1996a} criteria,
but would be identified in Distant Red Galaxy (DRG) samples \citep[e.g.,][]{franx2003a}
with $(\mJ-\mKs)=3.2$ and $(\mI-\mKs)=5.6$ (Vega) at $z\sim3$.
The quasar descendant would eventually obtain a $g'-r' = 1.24$ color 
(calculated assuming passive evolution to $z\sim0.2$) typical of 
brightest cluster galaxies in the MaxBCG cluster catalogs of the SDSS \citep[e.g.][]{koester2007a}.

The SEDs calculated for each of the three phases in the photometric evolution 
of the quasar host galaxy can be used to follow the trajectory of the system through
various color-color spaces. The magnitudes calculated according the method described
in \S \ref{section:telescope_response} can be compared with color criteria from the
literature to determine whether color-selected samples would find massive
quasar host galaxies.  In what follows, we examine whether the simulated galaxy 
satisfies common color criteria used in previous observational work.  For convenience,
the quasar host galaxy magnitudes in the salient observational filters at twenty 
redshifts $z=14-4.25$ are presented in Table \ref{table:seds}.

\small
\begin{deluxetable*}{lccccccccccccccc}
\tablewidth{0pt}
\tablecaption{\label{table:seds}Photometric Evolution (AB Magnitudes)
}
\tablehead{
\colhead{Redshift} & \colhead{F435W} & \colhead{F606W} &  \colhead{F775W} & \colhead{i'} & \colhead{z'} & \colhead{F850LP} & \colhead{F110W} & \colhead{F160W} & \colhead{$\mathrm{K}_{s}$} & \colhead{$[3.6\mu m]$} & \colhead{$[4.5\mu m]$} & \colhead{$[5.8\mu m]$} & \colhead{$[8\mu m]$}}
\startdata
14.00 & $>30$ & $>30$ & $>30$ & $>30$ & $>30$ & $>30$ & $>30$ & $>30$ & $26.11$ & $26.27$ & $26.39$ & $26.40$ & $26.40$\\
13.18 & $>30$ & $>30$ & $>30$ & $>30$ & $>30$ & $>30$ & $>30$ & $27.77$ & $25.68$ & $25.81$ & $25.89$ & $25.65$ & $25.57$\\
12.32 & $>30$ & $>30$ & $>30$ & $>30$ & $>30$ & $>30$ & $>30$ & $24.94$ & $23.97$ & $24.09$ & $24.16$ & $23.80$ & $23.81$\\
11.04 & $>30$ & $>30$ & $>30$ & $>30$ & $>30$ & $>30$ & $>30$ & $25.08$ & $24.70$ & $24.60$ & $24.25$ & $23.83$ & $23.83$\\
10.03 & $>30$ & $>30$ & $>30$ & $>30$ & $>30$ & $>30$ & $27.08$ & $23.61$ & $23.56$ & $23.47$ & $22.92$ & $22.79$ & $22.74$\\
9.600 & $>30$ & $>30$ & $>30$ & $>30$ & $>30$ & $>30$ & $24.01$ & $21.67$ & $21.75$ & $21.96$ & $21.59$ & $21.61$ & $21.61$\\
9.209 & $>30$ & $>30$ & $>30$ & $>30$ & $>30$ & $>30$ & $23.83$ & $21.98$ & $21.98$ & $22.03$ & $21.57$ & $21.57$ & $21.48$\\
8.555 & $>30$ & $>30$ & $>30$ & $>30$ & $29.85$ & $29.22$ & $21.69$ & $20.45$ & $20.49$ & $20.49$ & $20.25$ & $20.26$ & $20.11$\\
7.970 & $>30$ & $>30$ & $>30$ & $>30$ & $27.92$ & $27.84$ & $21.93$ & $20.92$ & $20.92$ & $20.30$ & $20.08$ & $20.05$ & $19.97$\\
7.467 & $>30$ & $>30$ & $>30$ & $>30$ & $27.14$ & $26.00$ & $22.35$ & $21.49$ & $21.41$ & $20.40$ & $20.28$ & $20.20$ & $20.12$\\
7.008 & $>30$ & $>30$ & $27.74$ & $27.45$ & $24.58$ & $24.04$ & $21.96$ & $21.32$ & $21.12$ & $20.07$ & $19.96$ & $19.81$ & $19.80$\\
6.607 & $>30$ & $>30$ & $25.62$ & $25.47$ & $22.69$ & $22.50$ & $21.31$ & $20.84$ & $20.55$ & $19.72$ & $19.56$ & $19.39$ & $19.52$\\
6.467 & $>30$ & $>30$ & $26.69$ & $26.55$ & $23.74$ & $23.57$ & $22.38$ & $21.84$ & $21.51$ & $20.18$ & $20.06$ & $19.99$ & $19.94$\\
6.300 & $>30$ & $29.90$ & $25.81$ & $25.71$ & $22.95$ & $22.87$ & $22.09$ & $21.66$ & $21.31$ & $20.15$ & $20.00$ & $19.90$ & $19.91$\\
5.993 & $>30$ & $27.99$ & $24.94$ & $24.75$ & $22.24$ & $22.22$ & $21.89$ & $21.58$ & $21.21$ & $20.16$ & $19.96$ & $19.87$ & $19.94$\\
5.563 & $>30$ & $26.68$ & $23.91$ & $23.76$ & $22.57$ & $22.55$ & $22.24$ & $21.96$ & $21.48$ & $20.25$ & $20.06$ & $20.00$ & $20.02$\\
5.007 & $>30$ & $26.22$ & $23.80$ & $23.69$ & $23.21$ & $23.19$ & $22.87$ & $22.40$ & $21.58$ & $20.23$ & $20.10$ & $20.03$ & $20.02$\\
4.491 & $>30$ & $25.28$ & $23.68$ & $23.65$ & $23.37$ & $23.35$ & $23.02$ & $22.26$ & $20.91$ & $20.07$ & $19.98$ & $19.91$ & $19.93$\\
4.250 & $29.51$ & $24.99$ & $23.79$ & $23.77$ & $23.49$ & $23.47$ & $23.12$ & $22.22$ & $20.77$ & $20.06$ & $19.96$ & $19.89$ & $19.90$
\enddata
\end{deluxetable*}

\begin{figure}
\figurenum{4}
\epsscale{1}
\plotone{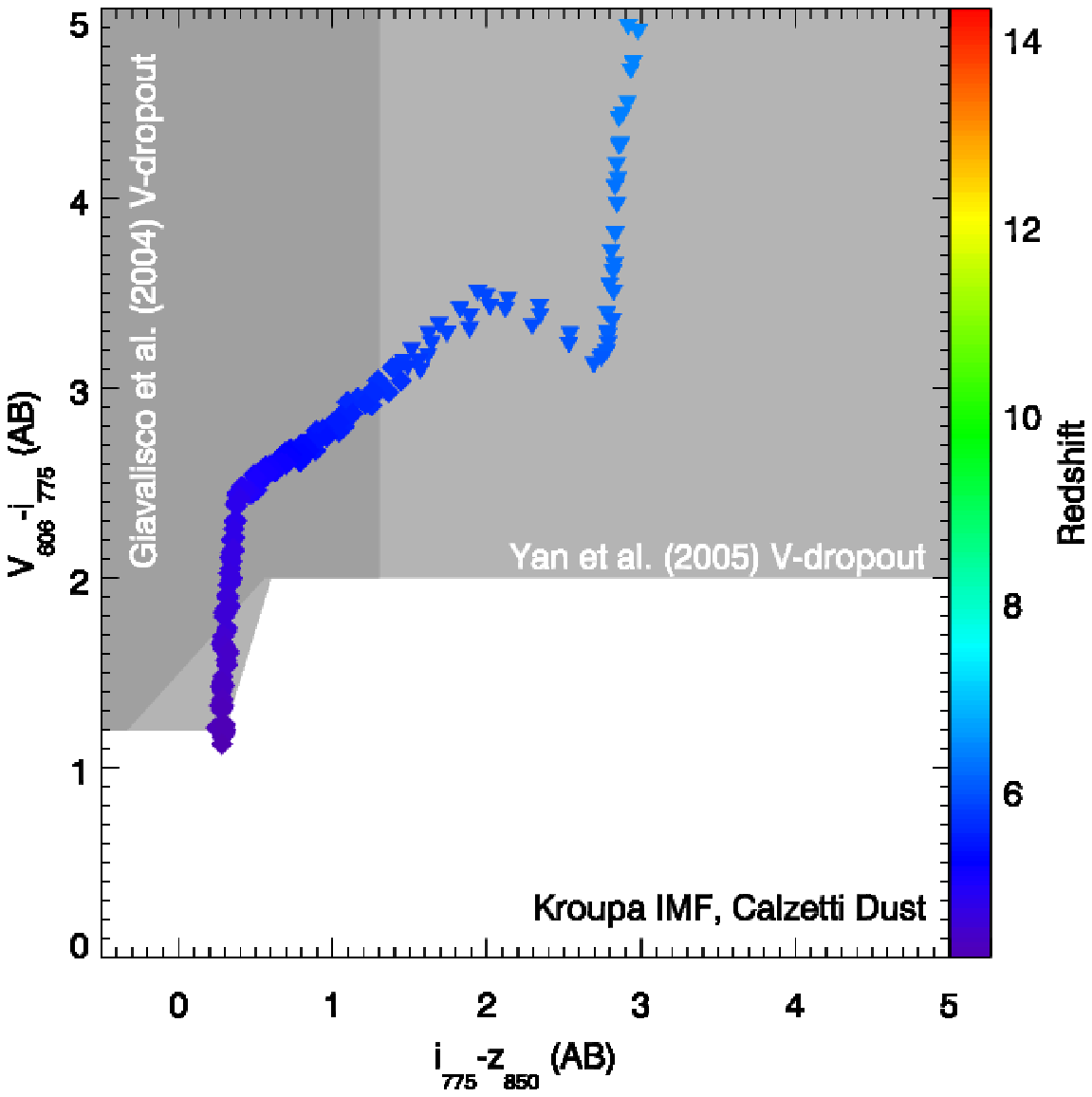}
\caption{\label{fig:v_dropout}
\small   Calculated $\mACSV-\mACSi$ and $\mACSi-\mACSz$ colors with 
redshift for a
simulated quasar host galaxy spectral energy distribution (SED).  Shown is the 
location of the SED in this color-color space when the system SED has
significant AGN activity (triangles, $5.7\lessim z \lessim6.7$) or is passively-evolving 
(diamonds, $z\lessim5.7$).  Also shown are the \cite{giavalisco2004b} GOODS and
\cite{yan2005a} HUDF
color selection criteria for galaxies at $z\sim5$.
The simulated galaxy satisfies these criteria during its passive evolution phase.
At higher redshifts, during the epoch of AGN activity,
the system will move into $\mACSi$-dropout
samples ($\mACSi-\mACSz\gtrsim 1.3$, see Fig. \ref{fig:i_dropout}).
}
\end{figure}

\subsection{$\mACSV$-dropout Selection}
\label{section:v_dropout}

The multi-color HST ACS observations in the GOODS fields allow for a variety of
color-color selection criteria to identify LBG samples.  The bluest color
Lyman-break dropout techniques for HST data that overlap with the simulation redshift
coverage are the  $\mACSV$-dropout selections at $z\sim5$.  
\cite{giavalisco2004b} presented a $\mACSV$-dropout selection for GOODS 
by the color criterion
\begin{eqnarray}
\label{equation:v_dropout_giavalisco}
\left\{\left[(\mACSV-\mACSi)>1.5 \right.\right.&+& \left.0.9 \times (\mACSi-\mACSz)\right] \nonumber\\
\OR\left[(\mACSV-\right.&\mACSi&) >    \left.\left. 2.0\right]\right\}  \nonumber\\
&\AND&\nonumber\\
\left\{(\mACSV-\right.&\mACSi&)  \geq \left.1.2\right\}\nonumber\\
&\AND&\nonumber\\
\left\{(\mACSi-\right.&\mACSz&)  \leq \left.1.3\right\},
\end{eqnarray}
\noindent
where $\OR$ and $\AND$ represent the logical ``OR'' and ``AND'' operations, and
used galaxies selected in this manner to estimate the global star formation rate
density at $z\sim5$.
A similar selection criterion, defined as
\begin{eqnarray}
\label{equation:v_dropout_yan}
\left\{\left[ (\mACSV-\mACSi)>1.2 \right] \right. &\AND& \left. \left[(\mACSi-\mACSz)\leq0.3\right]\right\} \nonumber\\
&\OR&\nonumber\\
\left\{\left[ (\mACSV-\mACSi)>0.4 \right. \right. &+&\left. 2.67(\mACSi-\mACSz)\right] \nonumber\\
\AND\left[ 0.3\leq(\mACSi\right.&-&\left.\left.\mACSz)\leq0.6\right]\right\} \nonumber\\
&\OR&\nonumber\\
\left\{\left[ (\mACSV-\mACSi)\geq2.0 \right] \right.&\AND& \left. \left[(\mACSi-\mACSz)>0.6\right] \right\},
\end{eqnarray}
\noindent
was used by \cite{yan2005a} to find $z\sim5$ galaxies in the HUDF for
examination with coincident \it Spitzer \rm IRAC observations.  The application
of stellar population synthesis modeling found that galaxies at 
$z\sim5$ in GOODS have typical stellar masses of $\Mstar\sim10^{10}\Msun$,
stellar ages of $t\sim1\Gyr$, low extinction, and a wide range of metallicities.

\begin{figure}
\figurenum{5}
\epsscale{1}
\plotone{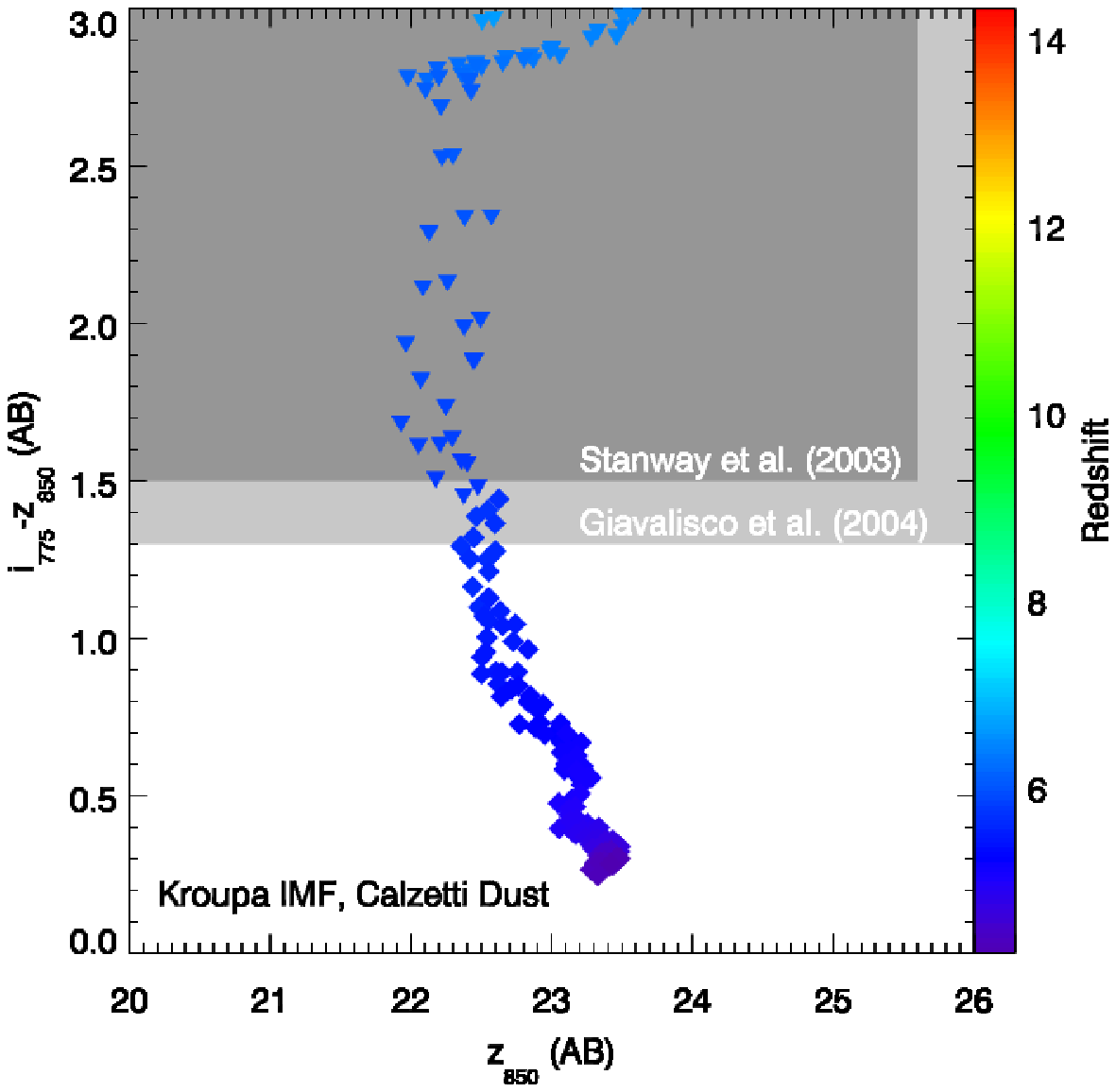}
\caption{\label{fig:i_dropout}
\small   Calculated $\mACSi-\mACSz$ color and $\mACSz$ magnitude with 
redshift for a
simulated quasar host galaxy spectral energy distribution (SED).  Shown is the 
location of the SED in this color-color space when the system SED is roughly
starburst-dominated (circles, $z\gtrsim6.7$), shows
significant AGN activity (triangles, $5.7\lessim z \lesssim6.7$) or is passively-evolving 
(diamonds, $z\lesssim5.7$).  Also shown are the \cite{giavalisco2004b} and
\cite{stanway2003a} GOODS
color selection criteria for galaxies at $z\sim6$.
The simulated galaxy satisfies these criteria during the epoch of AGN activity and,
owing to its activity and large stellar mass, is 
considerably brighter than the
candidate $z\sim6$ GOODS galaxies.
}
\end{figure}

Figure \ref{fig:v_dropout} shows the calculated 
$\mACSV-\mACSi$ and $\mACSi-\mACSz$ colors for the massive
host galaxy at redshifts $z>4$.  The system satisfies the
$\mACSV$-dropout color criterion over its entire 
passive-evolution phase at redshifts $z\lessim5.7$ when the
SMBH accretion and star formation rates are low and the
stellar component is passively aging.  In $\mACSV$-dropout
samples, the massive descendants of $z\sim6$ quasars could
appear as bright objects ($\mACSi\sim23.5$) with red $\mKs-\mIRACa$
colors (see Fig. \ref{fig:seds} or Table \ref{table:seds}).
At higher redshifts ($z\gtrsim5.7$), during an epoch of 
AGN activity,
the system will move into $\mACSi$- and $i'$-dropout
samples ($\mACSi-\mACSz\gtrsim 1.3$, see Fig. \ref{fig:i_dropout} 
and \S \ref{section:i_dropout}).

\subsection{$\mACSi$-dropout Selection}
\label{section:i_dropout}

Earlier in the evolution of the quasar host galaxy,
during its phase of significant AGN activity 
at times $5.7\lessim z\lessim6.7$, the IGM
absorption has moved into the $\fACSi$ filter
and the system becomes red in the $\mACSi-\mACSz$
color.  
To identify $\mACSi$-dropouts at $z\sim6$ 
in the GOODS ACS data,
\cite{giavalisco2004a} and \cite{dickinson2004a} 
used a simple criterion given
by
\begin{eqnarray}
\label{equation:i_dropout_giavalisco}
(\mACSi-\mACSz)\geq1.3.
\end{eqnarray}
\noindent
The GOODS data was also used to define a 
an $\mACSi$-dropout
sample by \cite{stanway2003a}, using 
similar criteria
\begin{eqnarray}
\label{equation:i_dropout_stanway}
&\left\{\mACSz\right.\left.<25.6\right\}&\nonumber\\
&\AND&\nonumber\\
&\left\{(\mACSi-\mACSz)\right.\geq\left.1.5\right\}.&
\end{eqnarray}
\noindent
\citep[see also][]{bunker2004a}.
Both
samples were used to estimate the global star 
formation rate density at $z\sim6$.

Figure \ref{fig:i_dropout} shows the $\mACSi-\mACSz$ color and $\mACSz$ magnitude 
of the quasar host galaxy during its evolution.
The simulated galaxy satisfies these criteria during its 
phase of AGN activity and,
owing to its activity and large stellar mass ($\Mstar=1.64\times10^{12}\Msun$ at $z=6.5$), is 
considerably brighter ($\Delta \mACSz \sim 3$) than
candidate $z\sim6$ GOODS galaxies ($\mACSz\sim25.5$, near the \cite{stanway2003a} magnitude cut).

\begin{figure}
\figurenum{6}
\epsscale{1}
\plotone{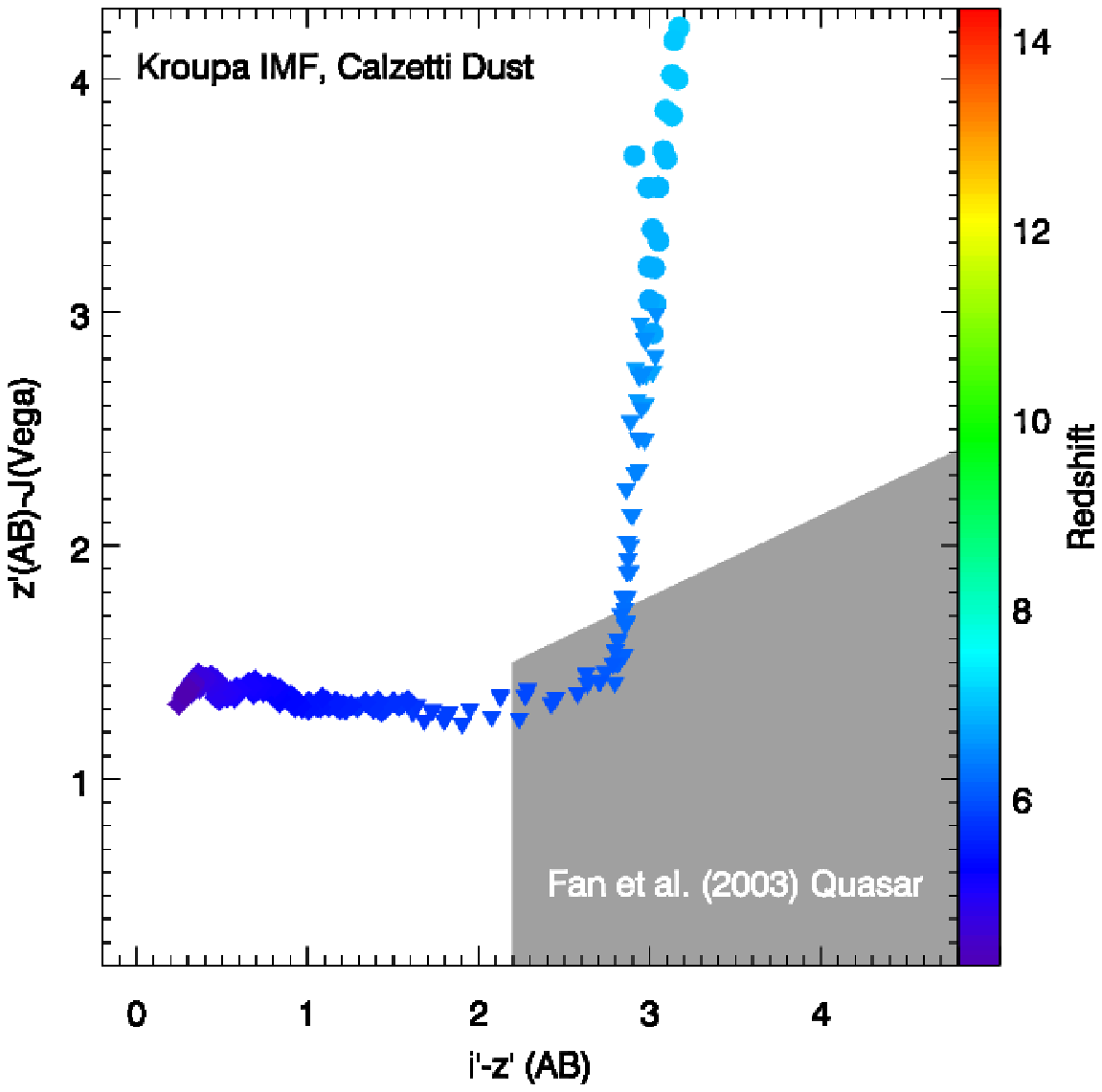}
\caption{\label{fig:sloan_quasars}
\small   Calculated $z_{\AB}'-J_{\Vega}$ and $i_{\AB}'-z_{\AB}'$ colors with 
redshift for a
simulated quasar host galaxy spectral energy distribution (SED).  Shown is the 
location of the SED in this color-color space when the system SED is roughly
starburst-dominated (circles, $z\gtrsim6.7$), has significant AGN activity
(triangles, $5.7\lesssim z \lesssim6.7$), or is passively-evolving 
(diamonds, $z\lessim5.7$).  Also shown is the SDSS
color selection criterion for $z\sim6$ quasars from \cite{fan2003a}.
The simulated galaxy satisfies the \cite{fan2003a} color criterion during its quasar
phase near $z\sim6$ and color-selected as a SDSS quasar candidate.  However, the 
system is 
$\sim2$ magnitudes fainter than the $z<20.2$ criterion used by 
\cite{fan2003a} to define their quasar sample. 
}
\end{figure}       

\begin{figure}
\figurenum{7}
\epsscale{1}
\plotone{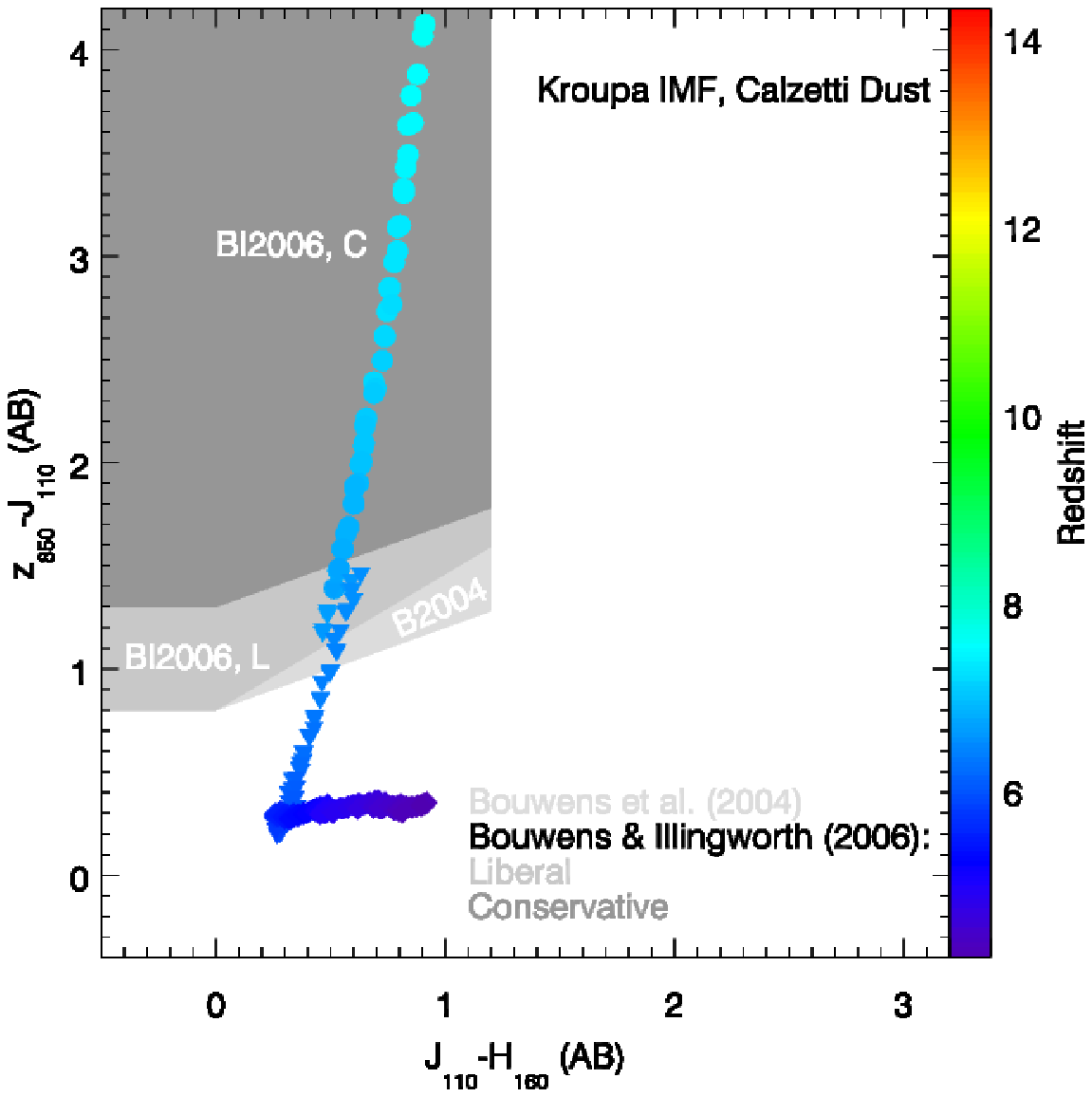}
\caption{\label{fig:z_dropout}
\small   Calculated $\mACSz-\mNICMOSJ$ and $\mNICMOSJ-\mNICMOSH$ colors with 
redshift for a
simulated quasar host galaxy spectral energy distribution (SED).  Shown is the 
location of the SED in this color-color space when the system SED is
starburst-dominated (circles, $z\gtrsim6.7$), shows significant
AGN activity (triangles, $5.7\lesssim z \lesssim6.7$), or is passively-evolving 
(diamonds, $z<5.7$).
Also shown are the \cite{bouwens2004a} (light gray), \cite{bouwens2006a}-``liberal''
(gray), and \cite{bouwens2006a}-``conservative'' (dark gray) 
color selection criteria for galaxies at $z\gtrsim7$ in NICMOS observations of
the Hubble Ultra Deep Field (HUDF).
The simulated $z\sim6$ quasar host galaxy satisfies these criteria primarily during its
starburst-dominated evolution phase at redshifts $z\sim7-8$, but owing to its extreme
star-formation rates and stellar mass it is typically $\sim6$
$\fNICMOSH$-magnitudes brighter than HUDF candidates.
}
\end{figure}       

\subsection{SDSS Quasar Selection}
\label{section:sloan_quasar}

The identification of quasars by multi-color selection criteria has been a common
practice for nearly twenty years \citep[e.g][]{warren1987a,irwin1991a},
and has proven especially fruitful for identifying distant ($z\sim6$) quasars 
in the SDSS 
\citep[e.g.,][]{fan2000a,fan2001a,fan2003a,fan2006a}.
The revised selection criterion used by \cite{fan2003a}, given by
\begin{eqnarray}
\label{equation:sdss_selection}
&\left\{z' \right.<\left. 20.2\right\}&\nonumber\\
&\AND&\nonumber\\
&\left\{(i'-z'\right.)\left.> 2.2\right\}&\nonumber\\
&\AND&\nonumber\\
&\left\{(z' - J_{\Vega})\right. < 1.5 +\left. 0.35(i'-z'-2.2)\right\},&
\end{eqnarray}
\noindent
incorporates $i'$-dropouts determined by the SDSS data with supplemental 
$J$-band photometry to distinguish from objects with red $z'-J$ colors, such
as cool stars.
Figure \ref{fig:sloan_quasars} shows the $z_{\AB}'-J_{\Vega}$ and $i_{\AB}'-z_{\AB}'$ colors 
calculated for the quasar host galaxy over its evolution, including the phase of strong AGN
activity at $z\sim6.5$.
The simulated galaxy satisfies the \cite{fan2003a} color criteria during its quasar
phase near $z\sim6.5$ and would be color-selected as a SDSS quasar candidate.  However, the 
system is 
$\sim2$ magnitudes fainter than the $z*<20.2$ criterion used by 
\cite{fan2003a} to define their quasar sample.  The quasar phase in 
this simulation ($z\sim6.5$) occurs near the beginning of 
a broader phase of accretion
and AGN activity ($7\gtrsim z \gtrsim 5.5$),
before the SMBH mass reaches its final mass.  Had the peak unobscured
SMBH accretion occurred slightly later (e.g., $z\sim5.8$ rather than $z\sim6.5$),
the system would have met the magnitude cut for the \cite{fan2003a} SDSS quasar
selection.  We note that at $z\gtrsim 6.3$ the system moves near the L-dwarf locus
in the $\mzprime-J$ vs. $\miprime-\mzprime$ color space, similar to the 
behavior of the quasar evolutionary track examined by \cite{fan2003a} to inform
their choice of color selection criteria.

\subsection{$\mACSz$-dropout Selection}
\label{section:z_droput}

The combined ACS and NICMOS observations of the HUDF and GOODS fields
have allowed for a search of $z\sim7-8$ galaxies by looking for $\mACSz$-dropouts.
\cite{yan2004b} found three $\mACSz$-dropout galaxies in the HUDF
but concluded that their red $\mNICMOSJ-\mNICMOSH$ colors made them unlikely
to reside at $z\sim7$. \cite{bouwens2004a} and \cite{bouwens2006a} used 
$\mACSz$-dropout criteria with a blue $\mNICMOSJ-\mNICMOSH$ color cut
to find $z\sim7$ galaxies more robustly.  These color-color criteria include
the selection
\begin{eqnarray}
\label{equation:z_dropout_bouwens04}
&\left\{\left(\mACSz-\mNICMOSJ\right)>0.8\right\}&\nonumber\\
&\AND\nonumber&\\
&\left\{\left(\mACSz-\mNICMOSJ\right)>0.8 + 0.66\left(\mNICMOSJ-\mNICMOSH\right)\right\}&\nonumber\\
&\AND\nonumber&\\
&\left\{\left(\mNICMOSJ-\mNICMOSH\right)<1.2\right\}&
\end{eqnarray}
\noindent
from \cite{bouwens2004a}, the ``liberal'' color selection
from \cite{bouwens2006a}
\begin{eqnarray}
\label{equation:z_dropout_bouwens06_l}
&\left\{\left(\mACSz-\mNICMOSJ\right)>0.8\right\}&\nonumber\\
&\AND\nonumber&\\
&\left\{\left(\mACSz-\mNICMOSJ\right)>0.8 + 0.4\left(\mNICMOSJ-\mNICMOSH\right)\right\}&\nonumber\\
&\AND\nonumber&\\
&\left\{\left(\mNICMOSJ-\mNICMOSH\right)<1.2\right\},&
\end{eqnarray}
\noindent
and the \cite{bouwens2006a} ``conservative'' selection
\begin{eqnarray}
\label{equation:z_dropout_bouwens06_c}
&\left\{\left(\mACSz-\mNICMOSJ\right)>1.3\right\}&\nonumber\\
&\AND\nonumber&\\
&\left\{\left(\mACSz-\mNICMOSJ\right)>1.3 + 0.4\left(\mNICMOSJ-\mNICMOSH\right)\right\}&\nonumber\\
&\AND\nonumber&\\
&\left\{\left(\mNICMOSJ-\mNICMOSH\right)<1.2\right\}.&
\end{eqnarray}
\noindent
These photometric selections find $<5$ high-redshift galaxies in GOODS, 
suggesting a rapid
decline in the density of luminous star-forming galaxies 
and a decrease of the characteristic luminosity of galaxies by $\approx1$ AB 
magnitude from $z\gtrsim6$ to $z\sim7-8$.

Figure \ref{fig:z_dropout} shows the $\mACSz-\mNICMOSJ$ and 
$\mNICMOSJ-\mNICMOSH$ colors calculated for the quasar host galaxy with redshift.
The lower-redshift epochs that cover the AGN activity and passive-evolution phases
occur below $z\sim6.7$ and largely do not satisfy the $\mACSz$-dropout color
criteria.  The starburst phase at $z\gtrsim7$, when the luminosity of the system
is dominated by the young stars generated by the enormous star formation rates $\SFR\sim10^{3}-10^{4}$, covers the entire $\mACSz-\mNICMOSJ$ color extent of the 
\cite{bouwens2006a} conservative selection.  The starbursting quasar progenitor
grows to $\Mstar\approx1.6\times10^{12}\Msun$ in stars with a mass-weighted stellar
age of $\left<t\right>=250\Myr$ at $z=7$, powering a remarkable luminosity
$\sim6$ $\fNICMOSH$-magnitudes brighter than $\mACSz$-dropout $z\sim7$ 
candidates in GOODS.  If $z\sim6$ quasars form in the manner 
suggested by the \cite{li2006a}
simulations, a $\mACSz$-dropout or similar selection criterion would be
well-matched to find quasar progenitors during the peak of their spheroid formation.

\begin{figure}
\figurenum{8}
\epsscale{1}
\plotone{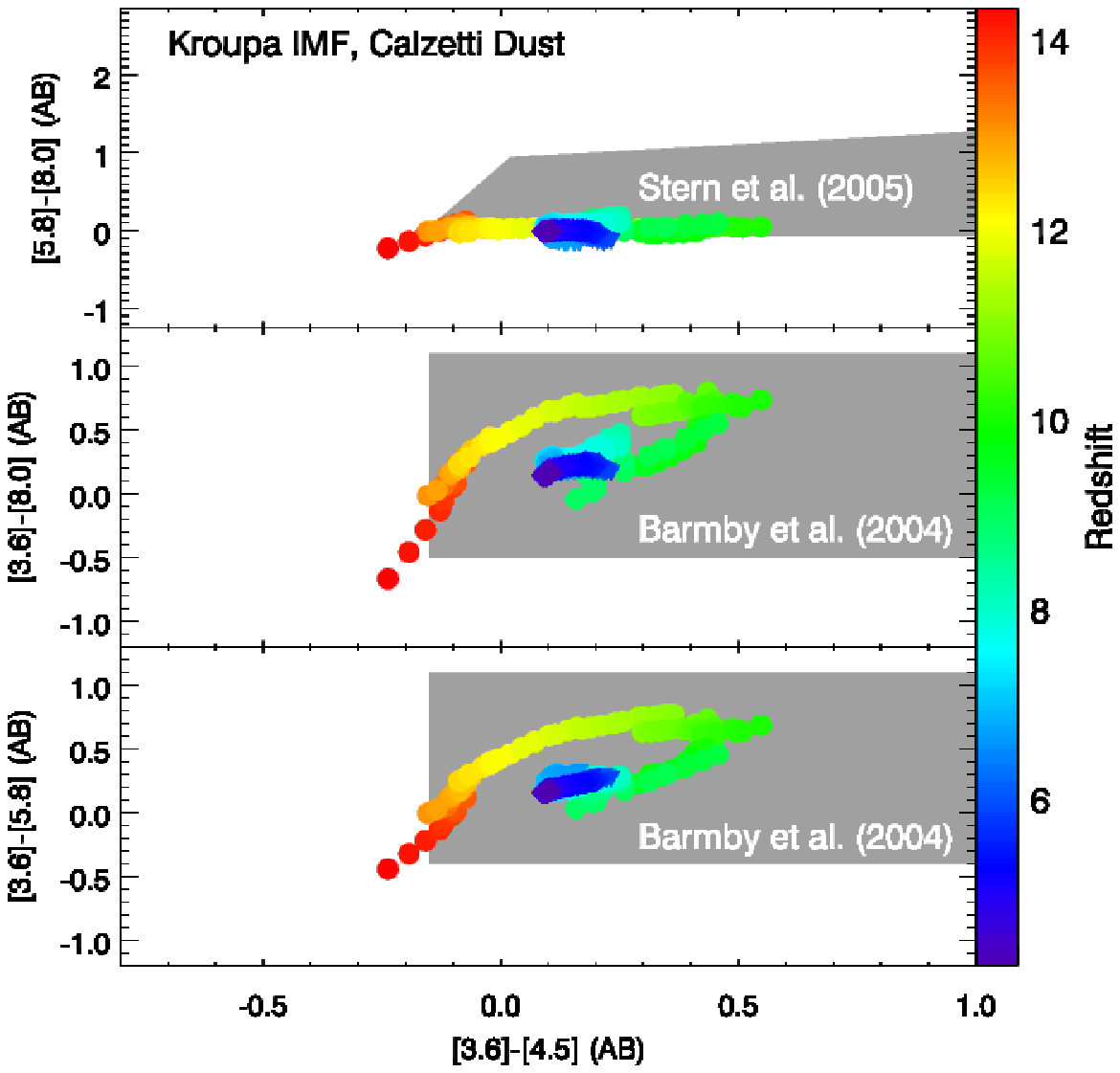}
\caption{\label{fig:IR_colors}
\small   Calculated Infrared Array Camera (IRAC) colors with 
redshift for a
simulated quasar host galaxy spectral energy distribution (SED).  Shown is the 
location of the SED in this color-color space when the system SED is
starburst-dominated (circles, $z\gtrsim6.7$), shows significant
AGN activity (triangles, $5.7\lesssim z \lesssim6.7$) or is passively-evolving 
(diamonds, $z<5.7$).  
Also shown are the AGES active galaxy IR-selection from \cite{stern2005a} 
and the
\cite{barmby2004a} IRAC 
color selection criteria for LBGs at $z\sim3$.
The simulated galaxy satisfies these criteria during most of its evolution and would
be detected in both the \cite{barmby2004a} IRAC observations 
and the IRAC Shallow Survey \citep{eisenhardt2004a}, but is $3-4$ AB magnitudes
fainter than the AGES sub-sample of the IRAC Shallow Survey.
}
\end{figure}       

\subsection{IRAC-based Selection}
\label{section:irac_selection}

In surveys at redshifts $z\lessim4$, the IR properties of both AGN
and galaxies have been used to design color-color selection criteria using 
IRAC bands.  \cite{stern2005b} showed that the $\mIRACa-\mIRACb$ color of
broad-line AGN at $1\lessim z \lessim4$ are typically redder than galaxies
at $z\lessim2$.  Their selection, given by,
\begin{eqnarray}
\label{equation:irac_stern}
&\left\{\left(\mIRACc-\mIRACd\right)>-0.07\right\}&\nonumber\\
&\AND&\nonumber\\
&\left\{\left(\mIRACa-\mIRACb\right)>-0.157\right.&\nonumber\\
&\left.+0.188\left(\mIRACc-\mIRACd\right)\right\}&\nonumber\\
&\AND&\nonumber\\
&\left\{\left(\mIRACa-\mIRACb\right)>-2.77\right.&\nonumber\\
&\left.+2.97\left(\mIRACc-\mIRACd\right)\right\}&
\end{eqnarray}
\noindent
identified 90\% of the spectroscopically identified type 1 AGN in the
AGN and Galaxy Evolution Survey (AGES).

\cite{barmby2004a} used template SEDs to
estimate the location of galaxies at $z\gtrsim2$ in the $\mIRACa-\mIRACc$ vs. 
$\mIRACa-\mIRACb$ and $\mIRACa-\mIRACd$ vs. $\mIRACa-\mIRACb$ color-color
spaces.  Comparing with a sample of confirmed $z\sim3$ LBG galaxies, they found that 
the majority of the LBGs resided in the IRAC color-color regions 
\begin{eqnarray}
\label{equation:irac_barmby_blue}
&\left\{\left( \mIRACa-\mIRACb \right)> -0.15 \right\}&\nonumber\\
&\AND&\nonumber\\
&\left\{ 1.1 > \left(\mIRACa-\mIRACc\right) > -0.4 \right\}&
\end{eqnarray}
\noindent
and
\begin{eqnarray}
\label{equation:irac_barmby_red}
&\left\{\left( \mIRACa-\mIRACb \right)> -0.15 \right\}&\nonumber\\
&\AND&\nonumber\\
&\left\{ 1.1 > \left(\mIRACa-\mIRACd\right) > -0.5 \right\}.&
\end{eqnarray}
\noindent
\cite{barmby2004a} combined these criteria with selections based on $\mRs$-band
magnitude.  At the redshifts we consider, the quasar host galaxy always has 
significant IGM attenuation of the $\mRs$-band and is significantly redder in
$\mRs-\mIRACa$ color than the galaxies examined by \cite{barmby2004a}.  We therefore
compare only with their IRAC color-color selections.

Figure \ref{fig:IR_colors} shows the calculated IRAC colors for
the simulated quasar host galaxy at redshifts $z>4$.   
The galaxy satisfies the \cite{barmby2004a} and \cite{stern2005a} 
criteria during most of its evolution.  The $\mIRACa-\mIRACb$
color begins blue at redshifts $z\gtrsim12$ when the star-forming
SED has its rest-frame near-UV and optical emission in the shortest-wavelength 
IRAC bands.  As the redshift decreases and the stellar populations age, the
$\mIRACa-\mIRACb$ color increases until the Balmer/4000$\Ang$~break passes through
these bands and the galaxy becomes very red in $\mIRACa-\mIRACb$.  At lower
redshifts, the system becomes relatively stable in its $\mIRACa-\mIRACb$ color
for the remainder of the simulation.
The system has a consistently flat $\mIRACc-\mIRACd$ color, and would allow
for its selection by the \cite{stern2005a} criteria.  However, the AGES
sub-sample of the IRAC Shallow Survey 
\citep{eisenhardt2004a} has a limiting magnitude
 $\sim3$ magnitudes fainter than the complete Shallow Survey in the $\mIRACa$-band,
and is too shallow to detect the simulated quasar host galaxy.  The
\cite{barmby2004a} color selections involve the $\mIRACa$-band and 
have considerably more evolution than redder bands since the 
$\mIRACa$-band transitions through the Balmer/4000$\Ang$~break 
(see \S\ref{section:balmer_break} below).  The large color-color
area of the \cite{barmby2004a} criterion would allow for the
quasar host galaxy to be selected over almost the entire duration
of the simulation (redshifts $z\lessim12$).
All three IRAC criteria examined would be
convenient
methods to define color selected samples that could include
quasar host galaxies at $z\gtrsim4$.

\begin{figure}
\figurenum{9}
\epsscale{1}
\plotone{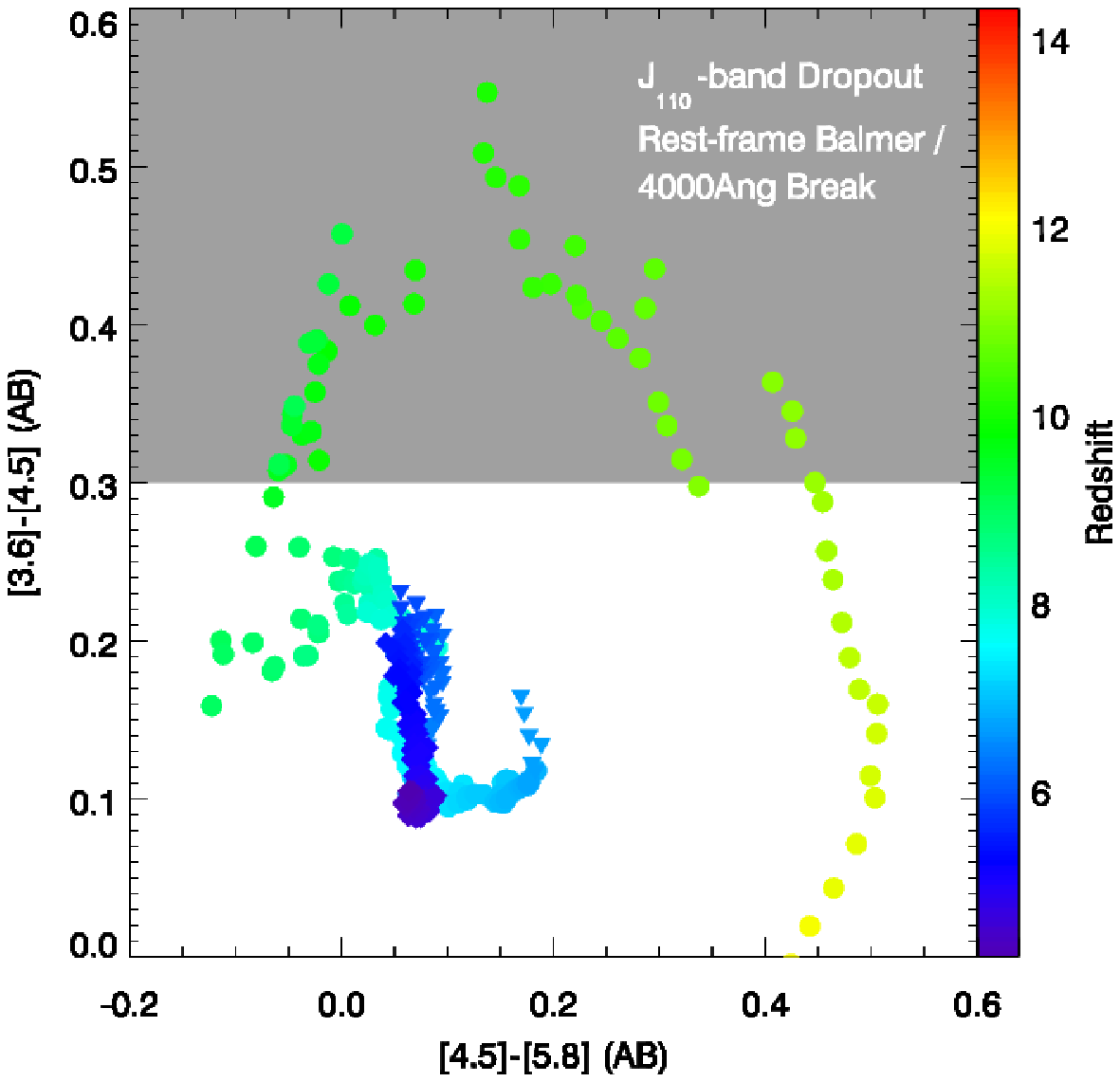}
\caption{\label{fig:highz_balmer}
\small   Calculated Infrared Array Camera (IRAC) $\mIRACa-\mIRACb$ vs. $\mIRACb-\mIRACc$
colors with 
redshift for a
simulated quasar host galaxy spectral energy distribution (SED).  Shown is the 
location of the SED in this color-color space when the system SED is
starburst-dominated (circles, $z\gtrsim6.7$), shows significant
AGN activity (triangles, $5.7\lesssim z \lessim 6.7$) or is passively-evolving 
(diamonds, $z<5.7$).  At $z\sim10$, the quasar progenitor has 
already been forming stars for $\gtrsim200\Myr$ and will display a rest-frame
Balmer/$4000\Ang$ spectral break. The break could be observed as a red $\mIRACa-\mIRACb$
color  in $\mNICMOSJ$-dropout samples, analogous to the $\mKs-\mIRACa$ spectral break
measured in $\mACSz$-band dropouts at lower redshifts \citep[e.g.,][]{labbe2006a}.
}
\end{figure}

\begin{figure}
\figurenum{10}
\epsscale{1}
\plotone{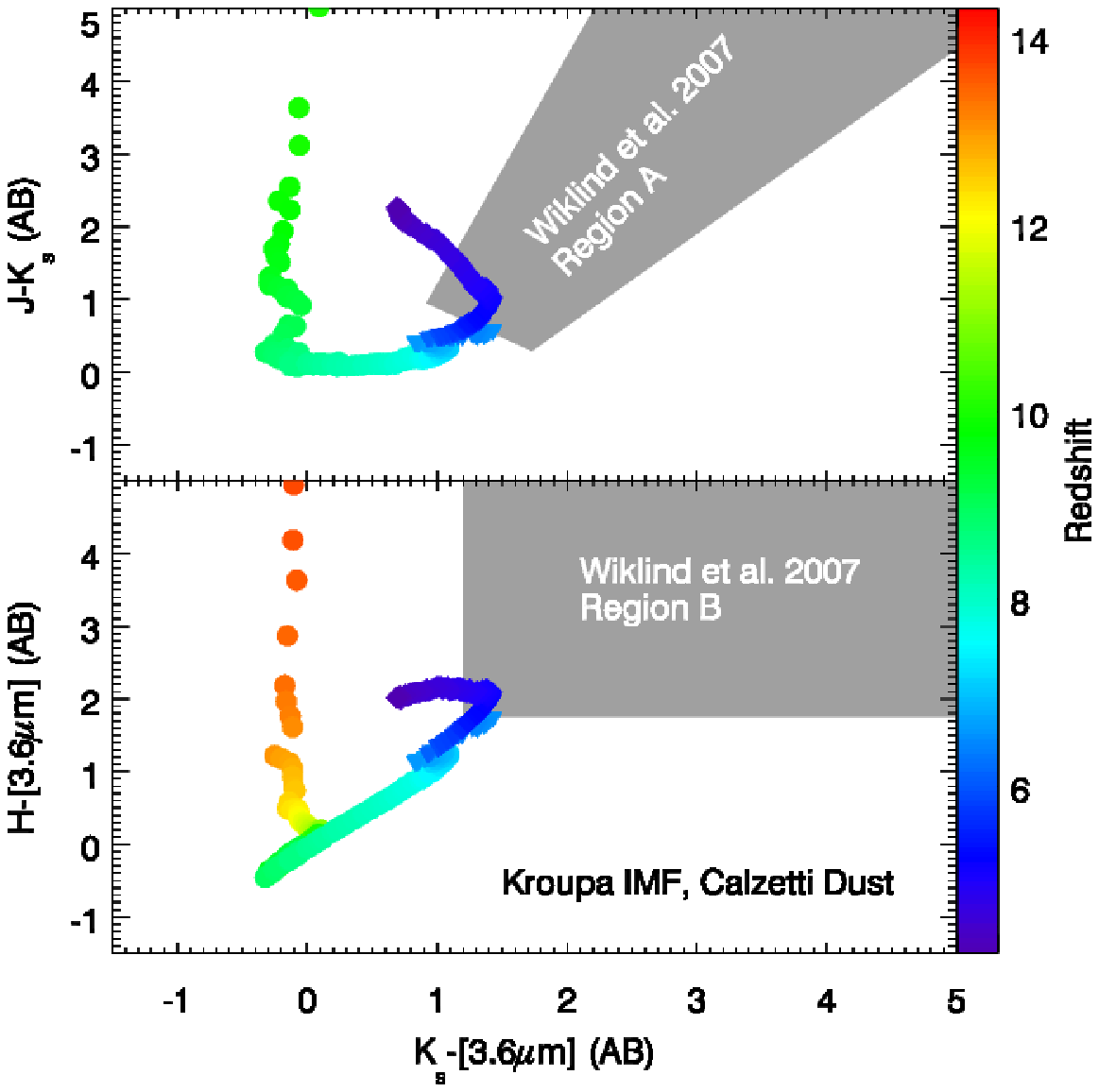}
\caption{\label{fig:wiklind}
\small   Calculated near-infrared $\mJ$, $\mH$, $\mKs$, and Infrared Array Camera (IRAC) $\mIRACa$
colors with 
redshift for a
simulated quasar host galaxy spectral energy distribution (SED).  Shown is the 
location of the SED in this color-color space when the system SED is
starburst-dominated (circles, $z\gtrsim6.7$), shows significant
AGN activity (triangles, $5.7\lessim z \lessim 6.7$) or is passively-evolving 
(diamonds, $z<5.7$).  Also shown are the ``Region A'' and ``Region B'' color selection
criteria from \cite{wiklind2007a,wiklind2007b} designed to identify post-starburst
systems at $z\sim5$ via the Balmer/$4000\AA$ break.  The quasar descendant satisfies the \cite{wiklind2007a,wiklind2007b}
criteria over the redshift range $4.8\lesssim z \lesssim5.5$, during the post-starburst,
passively-evolving phase of the system's evolution.  
The \cite{wiklind2007a,wiklind2007b} criteria are also briefly satisfied by the galaxy at $z\sim6.5$
during the height of the quasar activity.
}
\end{figure}

\subsection{High-Redshift 4000$\Ang$/Balmer Breaks}
\label{section:balmer_break}

The presence of 4000$\Ang$~or Balmer breaks in the observed SEDs of galaxies 
has been cited as evidence for evolved stellar populations in galaxies
at $z\gtrsim2$ \citep[e.g.][]{franx2003a}.  The calculated
quasar progenitor star formation history (Fig. \ref{fig:bhar_and_sfr})
suggests that a significant population of stars ($\Mstar\approx10^{11}\Msun$)
can form in such systems by $z\sim12$.  As these stars age they can 
contribute a small 4000$\Ang$ / Balmer spectral break in the galaxy SED, as is 
apparent in the $z\sim10$ SED in Figure \ref{fig:seds}, and may be
detected photometrically in the near-IR.

Figure \ref{fig:highz_balmer} shows the calculated 
IRAC $\mIRACa-\mIRACb$ vs. $\mIRACb-\mIRACc$
colors for the simulated quasar host galaxy at redshifts $z>4$.
The observable near-IR SED is very flat over the majority of the 
galaxy's evolution, remaining at $(\mIRACb-\mIRACc)<0.2$ and
$(\mIRACb-\mIRACc)<0.27$ at redshifts $z<9$.  At higher redshifts
$z\sim9-11$ when the SED would move into $\mNICMOSJ$-dropout 
samples,
the Balmer/4000$\Ang$~break moves into the $\mIRACa$-band and causes
the $\mIRACa-\mIRACb$ color to redder by $\sim0.2-0.3$ magnitudes.
As the Balmer/4000$\Ang$~break moves through the $\mIRACb$-band, the
evolutionary track of the SED reverses direction at higher redshifts
and the system
becomes substantially \it bluer \rm in $\mIRACa-\mIRACb$ as the filters begin to 
probe the region of the SED sensitive to the large star
formation rate (at $z\sim12$, see the upper left panel of
Fig. \ref{fig:seds}).  In contrast, the 
trajectory of 
local galaxy SEDs with mature stellar populations
in color-spaces sensitive to optical breaks
continue to redden as the SEDs are shifted to 
higher redshifts \citep[e.g., $J_{s}-K$ color at redshifts
$z>2.5$, see Figure 1 of][]{franx2003a}.  The
extremum of the $\mIRACa-\mIRACb$ at $z\sim9-11$
owing to the rest-frame optical break 
makes the color a useful indicator for the
presence of mature stars in $\mNICMOSJ$-dropout
samples.

A completely analogous
behavior can occur in lower redshift samples
in bluer bands.  SED fits to $\mACSz$-dropout
galaxies in the HUDF at $z\sim7$ show evidence for
a rest-frame optical break in their $K-\mIRACa$
colors \citep{labbe2006a}, as do $z\approx6$ galaxies
\citep{yan2005a,eyles2005a}, and possibly 
indicate an evolved stellar component.  Recently,
\cite{wiklind2007a,wiklind2007b} have proposed
Balmer break selection criteria for post-starburst
systems at $z\sim5$ using VLT/ISAAC near-IR $\mJ$,
$\mH$, and $\mKs$ and IRAC $\mIRACa$ band observations
of the GOODS fields.  Their
selection is defined by a union of objects in two regions:
a ``Region A'' in the $\mJ-\mKs$ vs. 
$\mH-\mIRACa$ color space, given by
\begin{eqnarray}
\label{equation:wiklind_region_A}
&\left\{\left( \mJ-\mKs \right)< -1.94 + 3.14\left( \mKs - \mIRACa \right )\right\}&\nonumber\\
&\AND&\nonumber\\
&\left\{\left( \mJ-\mKs \right)> -1.90 + 1.27\left( \mKs - \mIRACa \right )\right\}&\nonumber\\
&\AND&\nonumber\\
&\left\{\left( \mJ-\mKs \right)>  1.71 - 0.82\left( \mKs - \mIRACa \right )\right\},&
\end{eqnarray}
\noindent
and a 
``Region B'' in the $\mH-\mIRACa$ vs. 
$\mH-\mIRACa$ color space, given by
\begin{eqnarray}
\label{equation:wiklind_region_B}
&\left\{\left( \mH - \mIRACa \right)> 1.75 \right\}&\nonumber\\
&\AND&\nonumber\\
&\left\{\left(\mKs-\mIRACa\right) > 1.20 \right\}.&
\end{eqnarray}
\noindent
Figure \ref{fig:wiklind} shows the 
calculated near-infrared $\mJ$, $\mH$, $\mKs$, and IRAC $\mIRACa$
colors with redshift for the quasar host galaxy SED in relation to
Regions A and B from \cite{wiklind2007a,wiklind2007b}.
The quasar descendant satisfies the \cite{wiklind2007a,wiklind2007b}
criteria over the redshift range $4.8\lesssim z \lesssim5.5$, during the post-starburst,
passively-evolving phase of the system's evolution.  
The
comparison suggests that the \cite{wiklind2007a,wiklind2007b}
criteria would select the descendants of $z\sim6$ quasars
during the passively-evolving phase near redshift $z\sim5$.
While the \cite{wiklind2007a,wiklind2007b} selection was
designed to find massive ($\Mstar\sim10^{11}\Msun$) and evolved 
galaxies near $z\sim5$, the agreement may be somewhat surprising
given that the evolution of the simulated quasar host SED in the 
$\mJ$-, $\mH$-, $\mKs$-, and $\mIRACa$-bands is qualitatively
different than the evolving SED models used to design the
\cite{wiklind2007a,wiklind2007b} selection.  However, we note that
over the range of redshifts where the quasar host galaxy would
be detectable in the GOODS and ISAAC data the simulated system
resides in the locus of observed galaxies in the 
\cite{wiklind2007a,wiklind2007b} sample in the 
$(\mJ-\mKs)$--$(\mH-\mIRACa)$--$(\mKs-\mIRACa)$ color space.
We also note that the \cite{wiklind2007a,wiklind2007b} criteria 
are briefly satisfied by the galaxy at $z\sim6.5$
during the height of the quasar activity, but such bright sources
can be discriminated from inactive high-redshift galaxies through spectroscopic
observations.

\section{Observability}
\label{section:observability}

The comparisons between the redshift-dependent
photometric properties of massive
quasar host galaxies and color selection criteria in \S 
\ref{section:v_dropout}-\ref{section:balmer_break}
demonstrate that the SEDs of the most massive
high-redshift galaxies likely satisfy a variety of
existing photometric selection techniques.  However,
these high-redshift galaxies are extremely massive and 
therefore extraordinarily rare.  To determine 
whether existing or future galaxy surveys would
include such rare systems in their samples, the
observability of these massive galaxies must
be estimated.  A straightforward quantification
of the observability is to
calculate the fractional sky coverage and flux 
limit required to include and detect some defined
galaxy sample.  Below, we use the evolution of
the quasar host galaxy SED, the $\LCDM$ cosmological model,
and \cite{press1974a} theory to estimate these
quantities.

The comoving volume element $\dd V$ of a spherical
redshift shell of thickness $\dd z$ at comoving
distance $r(z)$ and redshift $z$ can be written
\begin{equation}
\label{equation:volume_element}
\dd V = 4\pi r^{2}(z) \frac{\dd r}{\dd z} \dd z
\end{equation}
\noindent
where
\begin{eqnarray}
\label{equation:comoving_distance}
&r(z) = \int_{0}^{z} \frac{\dd r}{\dd z'} \dd z'&\nonumber \\
&= \int_{0}^{z} \frac{c \dd z'}{H(z')},
\end{eqnarray}
\noindent
$c$ is the speed of light, the redshift-dependent Hubble parameter for a flat universe is
\begin{equation}
\label{equation:hubble_parameter}
H(z)=H_{0}\left[\OmegaM(1+z)^{3} + \OmegaL\right]^{1/2} \, ,
\end{equation}
\noindent
and $H_{0}$ is the Hubble constant today.
The number of galaxies $\dd N$ in this comoving volume element is then simply 
$\dd N =n(z)\dd V$, 
where $n(z)$ is the redshift-dependent comoving number density of galaxies that comprise
the sample.
For the form of $n(z)$, we use the mass function proposed by \cite{sheth1999a} to extend the
\cite{press1974a} theory of halo formation to model ellipsoidal 
collapse.  Assuming an overdensity 
threshold $\delta_{c}=1.686$ for halo collapse, we can define the 
quantity $\nu = \sqrt{a} \delta_{c}/[\sigma(M)D(z)]$ that relates the root-mean-squared 
fluctuations $\sigma(M)$ on a mass scale $M$, the linear growth factor 
\begin{equation} 
\label{equation:linear_growth_factor}
D(z) = D_{0} H(z) \int_{z}^{\infty} \frac{(1+z')\dd z'}{H^{3}(z')}
\end{equation}
with $D_{0}$ normalized such that $D(z=0)=1$, and $\delta_{c}$ to describe the rarity
of the density peak that collapsed to form the halo of mass $M$.  Here the constant $a=0.707$
was determined from cosmological N-body simulations by \cite{sheth1999a}. 
In terms of the peak rarity $\nu$, the comoving number density
of galaxies above some given mass $\Mlow$ is then
\begin{eqnarray}
\label{equation:comoving_number_density}
&n(\Mlow,z) = \int_{\ln \Mlow}^{\infty} \frac{\dd n}{\dd \ln m}\dd\ln m&\nonumber\\
&= \Omega_{m} \rho_{c}\int_{\ln \Mlow}^{\infty}\frac{A}{m\sigma} \sqrt{\frac{2}{\pi}}(1 + \nu^{-2p})\nu e^{-\frac{\nu^{2}}{2}}&\nonumber\\
&\times \frac{\dd \sigma}{\dd \ln m} \dd \ln m,&
\end{eqnarray}
\noindent
where the parameter $A=0.322$ is constrained such that 
$\int_{0}^{\infty}(\dd n /\dd \ln m)\dd\ln m = 1$ and $p=0.3$ is determined from fitting to
halo mass functions measured in cosmological simulations.  If we choose $\Mlow$ to correspond
to the redshift-dependent virial mass of a quasar host galaxy and assume that more massive halos will host
brighter galaxies, then $\Mlow$ defines a luminosity-selected sample of very massive galaxies 
at redshift
$z$ with comoving number density $n(z)$.

To calculate an actual number $N$ of galaxies, a redshift interval $\zlow\leq z \leq\zhigh$ must
be prescribed to 
define the comoving volume that hosts the sample as
\begin{eqnarray}
\label{equation:total_galaxy_number}
&N(\zlow,\zhigh,\Mlow) = \int n(\Mlow,z)\dd V&\nonumber\\
&=4\pi c\int_{\zlow}^{\zhigh}
\frac{n(\Mlow,z')r^{2}(z')\dd z'}{H(z')}.&
\end{eqnarray}
\noindent
For a given selection criterion, such as a Lyman-break dropout, $\zlow$ will correspond to the
redshift at which galaxy SEDs satisfy the color criteria.  At $\zlow$ the least massive galaxy
in the sample will have a magnitude $m_{\AB}$ that corresponds to the
brightest possible magnitude limit of a survey that could detect galaxies with mass 
$\Mlow$.  Typically, for dropout criteria this magnitude limit would apply to the redder of the
two bands (e.g., $\mACSi$-band for $\mACSV$-dropouts).  However, such a survey would have a 
very small comoving volume (since $\zlow=\zhigh$ in this case) and it is desirable to improve 
the magnitude limit to increase the comoving volume of the survey.  As the magnitude limit
of the survey is improved to a new $m_{\AB}'$, $\zhigh$ increases to the epoch when galaxies
with mass $\Mlow$ first became brighter than $m_{\AB}'$.  For a given photometric selection
which sets $\zlow$, and survey magnitude limit $m_{\AB}$, the fractional sky coverage needed to 
detect $N_{d}$ galaxies with mass
greater than $\Mlow$ at $\zlow$ is 
\begin{equation}
\label{equation:sky_coverage}
f = N_{d}/N(\zlow,\zhigh[m_{\AB}],\Mlow)
\end{equation}
\noindent
While this is conceptually straightforward, there are two immediate complications.  First,
the correspondence between the upper redshift limit $\zhigh$ and the magnitude limit $m_{\AB}$
depends on the redshift-dependent photometric properties of galaxies with mass $\Mlow$.
Second, galaxies with mass $\Mlow$ at epoch $\zlow$ are evolving and will have some lower
mass at higher redshifts.  Fortunately, the calculations of the redshift-dependent galaxy SED 
described in \S \ref{section:photometry} provide the desired $\zhigh-m_{\AB}$ correspondence.
For the mass accretion history the merger-tree from the \cite{li2006a} simulation is 
box-car averaged with window width of $\dd z=0.3$.
The resulting smooth mass accretion history increases rapidly from $z\sim14$ to $z\sim9$
owing to frequent mergers, and has a shape similar to the \cite{wechsler2002a} form for dark
matter halo mass accretion histories.  We note the observability calculation is insensitive to
the exact shape of the mass accretion history as long as the rapid decline in virial mass at
$z>9$ is reproduced.

\begin{figure*}
\figurenum{11}
\epsscale{0.8}
\plotone{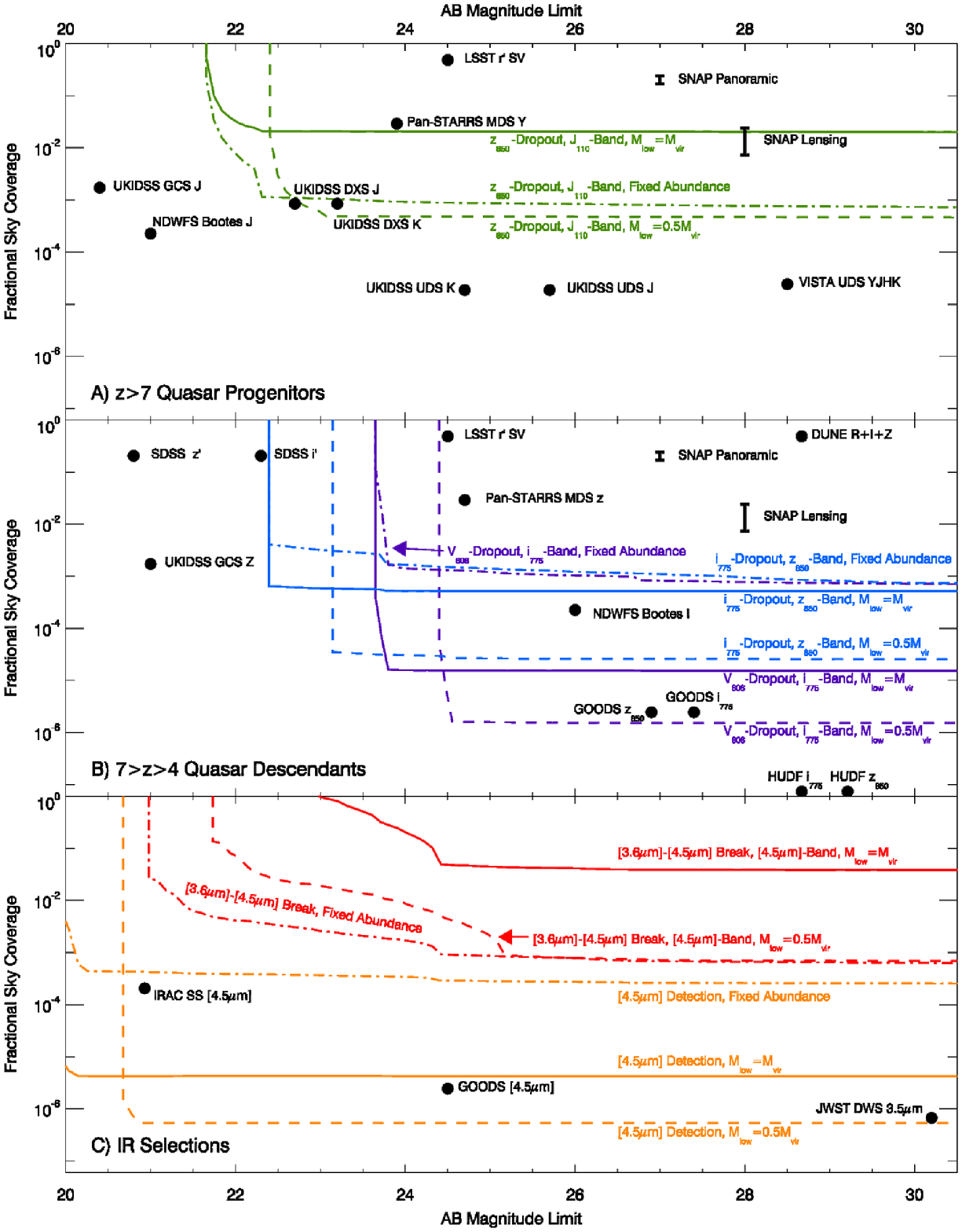}
\caption{\label{fig:surveys}
\tiny Estimated survey parameters required to find 
$z\gtrsim7$ quasar progenitors (upper panel) and
quasar descendants at redshifts $7\gtrsim z\gtrsim4$ (middle panel), or detect quasar host galaxies
in the near infrared (bottom panel).
Shown is the fractional sky coverage and minimum AB magnitude limit needed to build $\mACSV$-dropout (purple),
$\mACSi$-dropout (blue), $\mACSz$-dropout (green), and $\mIRACa-\mIRACb$ break (red)
samples that 
include a galaxy more massive than the virial mass $\Mvir$ (solid line) or $0.5\Mvir$ (dashed line) 
of the simulated $z\sim6$ quasar host from \cite{li2006a}.  
For comparison, the coverage needed to find objects with an abundance fixed at the number density of
halos more massive than the simulated galaxy at $z\sim6.5$ (dashed-dotted line) is also shown.
As the dropout selection moves to redder
bands and higher redshifts, the comoving volume and redshift interval over which massive galaxies satisfy
selection criteria decreases.  The comoving number density of massive galaxies, calculated using the 
\cite{sheth1999a} mass function, also declines rapidly at high-redshifts.  The combination these effects
requires large fractional sky coverage to find starbursting quasar progenitors at $z\gtrsim7$ (e.g. $\mACSz$-dropout
or $\mIRACa-\mIRACb$ break selections).  The circles show the parameters of the existing Hubble UDF ($\mACSi$- and
$\mACSz$-band),
GOODS ($\mACSi$-, $\mACSz$-, and $\mIRACb$-bands),
SDSS ($\miprime$- and $\mzprime$-bands), NOAO Wide Deep Field Survey ($I$- and $J$-bands),
UKIRT Deep Sky Survey ($Z$- and $J$- bands for the Galactic Clusters Survey, 
$J$- and $K$-bands for the Deep Extra Galactic and Ultradeep Surveys),
and IRAC Shallow Survey ($\mIRACb$-band) observations.
Future observations from the \it Dark Universe Explorer \rm ($R+I+Z$ band)
and the Ultra-deep \it Visible and Infrared Telescope for Astronomy \rm survey ($Y$-, $J$-,
$H$-, and $K$-bands) could detect $z\sim5-6$.
Even wider future surveys with redder sensitivity, such as the Pan-STARRS Medium Deep Survey
($z$- and $Y$-band), 
\it Large Synoptic Survey Telescope \rm observations
($\mrprime$ Single Visit), 
and the SNAP Panoramic and Lensing surveys (error bars) could find quasar 
progenitors at $z\gtrsim8$ if their two reddest-bands reach $\gtrsim22$ AB magnitude sensitivity.  
A \it James Webb Space Telescope \rm Deep-Wide Survey 
\citep{gardner2006a} 
would need to be considerably wider to find the $z>7$ starbursting 
progenitors of $z\sim6$ quasars.
Also
shown is the fractional sky coverage to detect $z\sim6$ quasar host galaxies at $z\gtrsim4$ in the $\mIRACb$-band
(orange lines), which only requires a small-area survey with a $\mIRACb\sim20.5$ magnitude limit.
}
\end{figure*}

The relation between fractional sky coverage and survey limiting magnitude is plotted in Figure 
\ref{fig:surveys} for a variety of color selection criteria explored in 
\ref{section:photometry}, including $\mACSV$-dropout 
(Eq. \ref{equation:v_dropout_giavalisco}, $\zlow\sim4.5$),
$\mACSi$-dropout (Eq. \ref{equation:i_dropout_giavalisco}, $\zlow\sim5.7$),
$\mACSz$-dropout (Eq. \ref{equation:z_dropout_bouwens06_c}, $\zlow\sim6.8$),
and $\mIRACa-\mIRACb>0.3$ break ($\zlow\sim9.1$) samples.
For these color selection criteria, the quasar host galaxy virial mass in the \cite{li2006a}
simulation increases from $\Mlow=5.22\times10^{12}h^{-1}\Msun$ at $\zlow=9.1$
to $\Mlow=8.01\times10^{12}h^{-1}\Msun$ at $\zlow\sim4.5$.
Beginning with the brightest possible magnitude limit, corresponding to the luminosity
of $\Mlow$-mass galaxies at $\zlow$, the necessary fractional sky coverage declines as 
the sample comoving volume increases with limiting magnitude until $\zhigh$ is reached.  
At larger (fainter) magnitudes, the fractional sky coverage becomes flat.
The width of the transition
region before the fractional sky coverage flattens is set by the redshift range over which the
luminosity declines.
For the $\mACSV$-, $\mACSi$-, and
$\mACSz$-dropout samples, IGM absorption sets $\zhigh$ to approximately 
the redshift when $\Lya$ moves into 
the $\mACSi$-, $\mACSz$-, and $\mNICMOSJ$-bands, respectively.  The Lyman break is dramatic at
such high redshifts and, when combined with the decrease in $\dd r/\dd z$ and $n(z)$ with 
redshift, causes the fractional sky coverage to flatten quickly beyond the brightest limiting magnitude.  
The larger transition width for the $\mACSz$-dropout selection relative to the bluer color
selections is set by the comparably large bandwidth of the $\fNICMOSJ$ filter.
For the
$\mIRACa-\mIRACb$-break selection, the fractional sky coverage artificially flattens at the 
$\mIRACb$-band 
magnitude of the system at $z\sim14$ ($\mIRACb\sim26$), where the simulation begins.  
Since these galaxies are bright ($\mIRACb<24$ to $z\sim12$), only 
a relatively small fractional sky coverage is needed to detect them.
However, given the 
decline of $\dd r/\dd z$ and $n(z)$ with redshift,
such a sample would be dominated by systems at $z\sim4$.

Given the rapid decline of the comoving number density $n$ with mass $\Mlow$ at fixed redshift, 
a useful comparison is the fractional sky coverage needed to observe systems that are less
massive than the \cite{li2006a} simulated galaxy by a factor of $\sim2$.  
Figure \ref{fig:surveys} also shows the require fractional sky coverage for these systems, 
assuming their photometric evolution is similar to massive galaxies and their luminosity scales
with their halo mass.  Such systems are clearly $\sim10\times$ more abundant than galaxies twice
their mass.  These systems would have stellar masses of $7\times10^{11}\Msun$, and 
SMBH masses $\MBH\sim10^{9}\Msun$, and could host quasars at $z\lessim6$.  
The virial mass of such galaxies is similar to that inferred from the clustering of
quasars at in the 2dF, SDSS, and other surveys \citep{porciani2004a,porciani2006a,wake2004a,croom2005a,coil2006a,shen2007a} and measurements
of the quasar proximity effect \citep{faucher_giguere2007a,kim2007a,guimaraes2007a}, and
are qualitatively similar to the system simulated by \cite{li2006a} and the large set of 
$z\sim6$ merger simulations performed by \cite{robertson2006a}. 
The detection of the host galaxies of such quasars
in Lyman-break dropout or Balmer-break samples would also provide interesting insights into
the formation of massive high-redshift galaxies 
\citep[for a possible sample of such systems at $z\sim5$, see][]{wiklind2007a,wiklind2007b}.

For comparison, we also show the necessary sky coverage to find objects with an abundance
fixed to the comoving number density ($n\sim4\times10^{-9} h^{-3}\Mpc^{-3}$) of halos with 
virial mass $\Mvir\gtrsim8\times10^{12} h^{-1}\Msun$
at $z\sim6.5$ if their luminosity evolution is similar to the simulated quasar host galaxy
(Fig. 11, dashed-dotted lines).  Correspondingly, such systems have an abundance similar 
to massive quasar host galaxies when they fall in $\mACSi$-dropout samples.  For higher-redshift
selections, this estimate of the massive galaxy abundance is more favorable for finding
the progenitors of $z\sim6$ quasars and is comparable to our $\Mlow=0.5\Mvir$ abundance 
estimates.

The estimated survey parameters for finding high-redshift massive galaxies can be
compared with the properties of existing and future photometric surveys to gauge
the likelihood for the most massive high-redshift galaxies to be found in actual
observational samples.  The largest existing survey is the SDSS, which has magnitude
limits of $\miprime\sim22.3$ and $\mzprime\sim20.8$ \citep{york2000a} over 8452 $\deg^{2}$~\footnote{http://www.sdss.org/status/}.  While the SDSS has sufficient area
to capture the most massive galaxies at redshifts as high as $z\gtrsim12$, the bright
magnitude limits and lack of infrared coverage makes the detection of extremely
massive high-redshift ($z>6$) galaxies in the SDSS implausible.  Of course, as has
been beautifully demonstrated, the rarer sub-population of massive galaxies in 
a bright quasar phase has been detected in $\sim1/2$ the total SDSS area \citep[e.g.,][]{fan2003a} as
the remarkable luminosity of the $z\sim6$ quasar sample and the large sky coverage
of the SDSS allows for the detection of tens of systems.

Compared with SDSS, deeper but smaller surveys have more advantageous parameters
for finding Lyman-break dropouts at redshifts $z\gtrsim4.5$.  For instance,
the Bo\"otes field of the National Optical Astronomy Observatory 
Wide Deep Field Survey \citep{jannuzi1999a} has an 9.3 $\deg^{2}$ field with
magnitude limits of $I=26$ and $J=21$.  While the $J$-band coverage is likely too
shallow to detect high-redshift quasar host galaxies, this area-magnitude combination should
be sufficient to find the most massive galaxies at $z\sim4.5-5$ in $V$-dropout
samples.  The accompanying IRAC Shallow Survey \citep{eisenhardt2004a}, with 8.5 $\deg^{2}$
coverage down to $\mIRACb\approx20.9$ would detect these systems as well as higher-redshift
$I$-dropouts in the NOAO NWDFS \citep[for galaxies with photometric redshifts $z\lessim2$, see][]{brodwin2006a}.  
The \it United Kingdom Infrared Telescope \rm Deep Sky Survey \citep[UKIDSS,][]{warren2002a,warren2007a} 
includes a Galaxy Cluster Survey (GCS, $70.9\deg^{2}$ area
to $Z\approx21$ and $J\approx20.4$), a Deep Extragalactic Survey (DXS, $35\deg^{2}$ area
to $J\approx23.2$ and $K=22.7$), and an Ultradeep Survey 
\citep[UDS, $0.77\deg^{2}$ area to $J\approx25.7$ and $K\approx24.7$, see][]{lawrence2007a}.
While the UKIDSS GCS is too shallow to find the very massive galaxies we model, the UKIDSS DXS
and UDS could detect such systems in the $J$- and $K$-bands at $z\gtrsim4$.  The UKIDSS DXS
may even have enough area to detect massive quasar progenitors at $z\gtrsim7$ \citep[see also][]{warren2002a}, 
which would be
an exciting prospect, but the lack of deep coverage at shorter wavelengths
in the DXS may limit its ability
to identify such systems as Lyman-break dropouts.
A proposed UDS with the Visible and Infrared Survey 
Telescope for Astronomy (VISTA) could reach 5-$\sigma$ AB limiting magnitudes
of $\mY=26.7$, $\mJ=26.6$, $\mH=26.1$, and $\Ks=25.6$ over a $\sim1 \deg^{2}$ field
(M. Franx, private communication).
While the proposed VISTA observations would be significantly deeper than the UKDISS UDS
or the NDWFS, the size of the field could limit the survey to finding 
$z\sim6$ quasar host galaxies at $z\lessim5$.

The GOODS observations that cover 365 arcmin$^{2}$ reach
depths of $\mACSi\approx27.4$ and $\mACSz\approx26.9$ with ACS \citep{giavalisco2004a}
and $\mIRACb=24.5$ with IRAC \citep{dickinson2003a}.  The GOODS $\mACSV$-dropout sample has
the necessary sky coverage to include a $\Mstar\sim5\times10^{11}\Msun$ stellar mass galaxy at
$z\sim4-5$ with $\mACSi\sim25$ and detections in all redder bands, though we are not aware
of any such systems in the current GOODS data.  Less massive ($\Mstar\sim10^{11}\Msun$) systems
are much more abundant and our estimate would predict a handful of such $z\sim5$ systems in GOODS,
which may have already been detected \citep[e.g.,][]{stark2006a}.
The HUDF data, with an 11 arcmin$^{2}$ area
and magnitude limits of $\mACSi=29.21$ and $\mACSz=28.67$ \citep{beckwith2006a}, is simply
too narrow to include objects as rare as $z\sim6$ quasar host galaxies.  We note briefly that these
simulations suggest that to detect $\Mstar\sim5\times10^{11}\Msun$ galaxies in $\mNICMOSJ$-dropout
samples at $z\gtrsim7$ 
\citep[e.g.][ see also \cite{chen2004a} and \cite{dunlop2006a}]{mobasher2005a}, a survey 3-4 orders
of magnitude larger than the HUDF would be necessary.

Future surveys with wide area coverage and deep imaging have a substantially better opportunity
to detect the starbursting progenitors of $z\sim6$ quasars.  The Pan-STARRS project is planning a
$1200 \deg^{2}$ Medium Deep Survey with magnitude limits of 
$z_{0.89\mum}=24.7$ and $Y_{1.02\mum}=23.9$\footnote{http://pan-starrs.ifa.hawaii.edu/}, which could
find massive quasar progenitors in $z$-band dropout samples at $z\gtrsim7$.  The \it Large Synoptic
Survey Telescope \rm would make short work of finding the most massive high-redshift
galaxies if its $Y$-band Single Visit Depth is comparable to its target 
$r=24.5$ depth\footnote{http://www.lsst.org/Science/docs/SRD\_summary.pdf}.
The \it Supernovae Acceleration Probe \rm (SNAP) is planning a weak 
lensing survey of $300-1000\deg^{2}$ 
to $m_{\AB}=28$ and a wide-area ($7000-10000\deg^{2}$) survey with a depth of 
$m_{\AB}=27$~\footnote{http://snap.lbl.gov}.  If a near-IR HgCdTe device flies with the mission
and reaches similar sensitivities down to $\lambda\sim1.5\mum$ over the same area, SNAP could
detect the progenitors of $z\sim6$ quasars at redshifts approaching $z\sim10$.  
The \it Dark Universe Explorer \rm mission will survey $20,000 \deg^2$ with a composite 
$R+I+Z$-band in the red optical and will reach an effective $I\sim24.5$ limiting 
magnitude \citep{refregier2006a}, which should detect massive spheroids at $z\sim5$.
The 
\it James Webb Space Telescope \rm easily has enough sensitivity to detect massive quasar progenitors
out to $z\gtrsim15$ (with stellar masses $\Mstar\lessim10^{10}\Msun$), 
but currently suggested observations, such as a Deep-Wide Survey with
a 100 arcmin$^{2}$ area and limiting magnitude of $m_{\AB}\approx30.2$ at $\lambda=3.5\mum$ 
\citep{gardner2006a}, would be too narrow to find such rare objects at the highest redshifts.

\section{Discussion}
\label{section:discussion}

The discovery of very massive high-redshift galaxies would provide an
interesting new perspective on the structure formation process.  While
the larger abundance of less-massive galaxies allow the 
redshift evolution of such systems to be researched by characterizing 
the statistical properties of populations of objects, the rarity of
the most massive galaxies will likely limit our ability to draw 
inferences about their formation from population studies.  If the
most massive galaxies at the current epoch were a largely heterogeneous
population, their rarity could prove a significant limitation in
unravelling the various modes for their formation.

Fortunately, the most massive galaxies today are a roughly
homogeneous population.  Studies of the color 
bimodality of galaxies
\citep[e.g.,][]{strateva2001a,bell2004a,baldry2004a}
show that the most luminous galaxies are almost uniformly
red ($\mgprime-\mrprime\gtrsim1.2$).  Such galaxies
populate the center of galaxy clusters and their
properties are roughly uniform between clusters at the
same epoch \citep[e.g.,][]{bower1992a,gladders2000a}.
Importantly, these massive galaxies have been demonstrated
to contain supermassive black holes with mass 
$\MBH\gtrsim10^{9}\Msun$
\citep[i.e. M87,][]{harms1994a,macchetto1997a}.
The large SMBH masses directly connect the population of 
the most massive galaxies at low redshift to the luminous
quasars at $z\sim6$, since $10^{9}\Msun$ SMBHs are the
most credible engine to power $z\sim6$ quasars, and suggest
that the most luminous high-redshift quasars evolve into
a roughly homogeneous population at the current epoch.  Our
calculations support this possibility, under the condition
that the most massive galaxies undergo little star formation
and their stellar populations evolve passively from 
$z\sim4$ to the present.  Observed
cluster ellipticals have stellar formation epochs of at least
$z\gtrsim2$ \citep[e.g.,][]{van_dokkum2001a,gebhardt2003a}.
We note that this picture does not preclude future major
mergers between spheroids at lower redshifts
\citep[e.g.,][]{van_dokkum2005a,bell2006a}, which may have
some relation to core-cusp bimodality 
\citep[e.g.,][Kormendy et al., in preparation, Krause et al., in preparation, though see \cite{ferrarese2006a}]{faber1997a,lauer2006a} that is
 the primary heterogeneity in massive galaxies.  Even so, this bimodality
exists primarily in systems with $M_{V}>-22$, with more 
luminous systems mostly displaying surface brightness cores.

The connection between the most massive galaxies in the
present epoch and quasars at $z\sim6$, forged by their
shared SMBH masses, and the uniformity of the most massive 
local galaxies permits an argument for a single evolutionary
scenario for the formation of the most massive galaxies
from very high-redshifts to the present day.  The appeal
of a single mode of formation for the most massive galaxies
stems primarily from the difficulty in forming luminous quasars 
at high-redshifts with $\gtrsim10^{9}\Msun$ SMBHs, the problem 
popularized by \cite{efstathiou1988b}.  Finding a robust 
way of growing SMBHs quickly in the limited time available before
$z\sim6$ is problematic, but the success of the \cite{li2006a}
simulation of $z\sim6$ quasar in producing a SMBH with a mass
of $\MBH>10^{9}\Msun$ has provided some hope that high-redshift
quasars can be explained naturally in the context of the
formation of rare density peaks in our $\LCDM$ cosmology.  
The calculations performed in this paper provide a detailed
characterization of the observable ramifications of this
scenario, the foremost being the possible detection of the 
star-bursting progenitors of $z\sim6$ quasars at higher
redshifts ($z\gtrsim6$) with massive stellar populations
($\Mstar\sim10^{11.5-12.0}$) in wide-area, Lyman-break dropout
samples or through wide-area IR searches for systems with
rest-frame optical breaks at $z\sim9$ and stellar masses of
$\Mstar\sim10^{11}\Msun$.

\section{Conclusions}

\label{section:conclusions}

Combining hydrodynamical simulations of the hierarchical formation of
a $z\sim6$ quasar \citep{li2006a}, stellar population synthesis models
\citep{leitherer1999a,vazquez2005a}, AGN spectral templates 
\citep{vanden_berk2001a,marconi2004a}, models for the 
wavelength-dependent attenuation owing to interstellar and intergalactic
absorption \citep{calzetti1994a,calzetti2000a,madau1995a}, and the
transmissivity of telescopes, filters, and detectors, the 
photometric properties of a massive $z\sim6$ quasar host galaxy are calculated
at redshifts $z\sim4-14$.  The photometric properties of the quasar
host galaxy reflect three main evolutionary phases in its formation;
a starburst-dominated phase at $z\gtrsim6.7$, an AGN phase at 
$5.7\lesssim z\lessim6.7$ that includes optical quasar activity at 
$z\sim6.5$, and a passive-evolution phase at $z\lessim5.7$.
The photometric properties of the system
are then compared with color-selection techniques for finding 
high-redshift galaxies.  The main findings of these calculations
follow.

\begin{itemize}

\item At very high-redshifts $z\sim6.7-14$,
multiple gas-rich mergers gives rise to star formation rates of
$\SFR\sim10^{3}-10^{4}$, building a stellar spheroid of $>10^{12}\Msun$
by $z\sim7$ \citep{li2006a}.  Our calculations show that these star formation rates
naturally give rise to a photometric starburst phase with strong rest-frame UV
and blue optical emission that satisfy HST ACS/NICMOS 
Lyman-break criteria for galaxies at $z>7$ 
\citep[e.g.][]{bouwens2004a,bouwens2006a}.
Over this epoch, the near-infrared luminosity of the system peaks
at $z\sim8.5$ with $\mNICMOSJ\sim21.7$, $\mNICMOSH\sim20.5$, and
$\mKs\sim20.5$ in AB magnitudes.

\item During the phase of AGN activity ($5.7\lessim z \lessim6.7$), IGM
absorption blue-ward of $\Lya$ causes the system to satisfy both $\fACSi$-band dropout
criteria for LBGs in GOODS \citep[e.g.][]{stanway2003a,giavalisco2004b}
and the SDSS $i'$-band dropout selection for high-redshift quasars
\citep[e.g.][]{fan2003a}.  At higher redshifts IGM absorption begins to
attenuate the $z'$-band flux and push the quasar host galaxy towards the
locus of L-dwarfs in the $z'-J_{\Vega}$ vs. $i'-z'$ color-color space.  During this
phase, the quasar
host galaxy is usually $\sim2$ $z'$-band magnitudes fainter than the 
\cite{fan2003a} sample selection flux limit.

\item At lower redshifts ($z\lessim5.7$) after the starburst and AGN activity 
has declined, the quasar host galaxy 
satisfies $\mACSV$-dropout criteria for galaxies at $z\sim5$ 
\citep[e.g.,][]{giavalisco2004a,yan2005a}.  The quasar descendant
also satisfies Balmer break selection techniques designed to identify
massive and evolved galaxies at $z\sim5$ \citep[e.g.,][]{wiklind2007a,wiklind2007b}.

\item Throughout the photometric evolution of the quasar host galaxy, the system
satisfies \it Spitzer \rm IRAC color-color selections for AGN \citep{stern2005a} and
high-redshift galaxies \citep{barmby2004a}.

\item The area and depth necessary to find the most massive high-redshift 
galaxies in photometric surveys is estimated from our photometric modeling and the
redshift-dependent abundance of dark matter halos \citep{press1974a,sheth1999a}.
Existing surveys, such as the IRAC Shallow Survey or NOAO Deep-Wide Field Survey, may
be large enough to find the massive $z\sim4$ descendants of $z\sim6$ quasars.
Surveys of significant fractions of the sky ($\gtrsim0.001$) in the $\mACSz$ and 
$\mNICMOSJ$ or similar bands, such as the UKIRT Deep Sky Survey, 
will be necessary to find the starbursting progenitors 
of $z\sim6$ at high-redshifts ($z>7$).  
Future surveys, such as Pan-STARRS, the 
\it Large Synoptic Survey Telescope\rm, or the 
\it Supernovae Acceleration Probe\rm, should have
the necessary area and depth to find the most massive high-redshift galaxies.

\item If future surveys
contain a $\mNICMOSJ$-dropout massive quasar progenitor at $z\sim10$, evidence
for star formation at $z\gtrsim12$ in the form of a 
Balmer/4000$\Ang$ break could be detected 
in the $\mIRACa-\mIRACb$ vs. $\mIRACb-\mIRACc$ color-color space.
\end{itemize}

The remarkable luminosity and early-formation time of the massive progenitor galaxies 
of $z\sim6$ quasars could enable their detection at very high-redshifts $z\gtrsim7$.
The detection of these massive galaxies would help settle outstanding questions about how
the rarest structures form in the $\LCDM$ cosmology and what evolutionary processes
give rise to $z\sim6$ quasars.  As observations must ultimately determine the existence and
properties of the most massive high-redshift galaxies, we eagerly anticipate 
future efforts to detect and characterize these extraordinary quasar progenitors.

\acknowledgements
BER gratefully acknowledges support from a Spitzer Fellowship,
through a NASA grant administered by the Spitzer Science Center, and
thanks Hsiao-Wen Chen, Tiziana Di Matteo, Volker Springel, Marijn Franx,
and Andrew Zentner for helpful comments.  BER also thanks Dr. Wiklind
for early access to his results before publication.
This work was supported in part by NSF grants ACI 96-19019, AST
00-71019, AST 02-06299, and AST 03-07690, and NASA ATP grants
NAG5-12140, NAG5-13292, and NAG5-13381.  The simulations were
performed at the Center for Parallel Astrophysical Computing at
Harvard-Smithsonian Center for Astrophysics.

 \end{document}